\documentclass{aa}  

\usepackage{graphicx}
\usepackage{txfonts}
\usepackage{bm}
\usepackage{xcolor}
\usepackage{hyperref}
\hypersetup{
    colorlinks=true,
    linkcolor=blue,
    citecolor=blue,
    urlcolor=cyan,
    }

\begin{document}

   \title{Impact of anti-solar differential rotation in mean-field solar-type dynamos}
   \subtitle{Exploring possible magnetic cycles in slowly rotating stars}

   \author{Q. Noraz\inst{1},
          A.S. Brun\inst{1},
          A. Strugarek\inst{1}\and
          G. Depambour\inst{1,2}
          }

   \institute{D\'epartement d'Astrophysique/AIM, CEA/IRFU, CNRS/INSU, Univ. Paris-Saclay, Univ. de Paris, 91191 Gif-sur-Yvette, France\\
              \email{quentin.noraz@cea.fr, allan-sacha.brun@cea.fr, antoine.strugarek@cea.fr}\\
              \and
              CentraleSupélec, Université Paris-Saclay, 91191 Gif-sur-Yvette, France\\
             }

   \date{Received 4 August, 2021; accepted 15 November, 2021}

 
  \abstract
   {Over the course of their lifetimes, the rotation of solar-type stars goes through different phases. Once they reach the zero-age main sequence (ZAMS), their global rotation rate decreases during the main sequence until at least the solar age, approximately following the empirical Skumanich’s law and enabling gyrochronology. Older solar-type stars might then reach a point of transition when they stop braking, according to recent results of asteroseismology. Additionally, recent 3D numerical simulations of solar-type stars show that different regimes of differential rotation can be characterized with the Rossby number. In particular, anti-solar differential rotation (fast poles, slow equator) may exist for high Rossby number (slow rotators). If this regime occurs during the main sequence and, in general, for slow rotators, we may consider how magnetic generation through the dynamo process might be impacted. In particular, we consider whether slowly rotating stars are indeed subject to magnetic cycles.}
   {We aim to understand the magnetic field generation of solar-type stars possessing an anti-solar differential rotation and we focus on the possible existence of magnetic cycles in such stars.}
   {We modeled mean-field kinematic dynamos in solar (fast equator, slow poles) and anti-solar (slow equator, fast poles) differential rotation, using the STELEM code. We consider two types of mean field dynamo mechanisms along with the $\Omega$-effect: the standard $\alpha$-effect distributed at various locations in the convective envelope and the Babcock-Leighton effect.}
   {We find that kinematic $\alpha\Omega$ dynamos allow for the presence of magnetic cycles and global polarity reversals for both rotation regimes, but only if the $\alpha$-effect is saddled on the tachocline. If it is distributed in the convection zone, solar-type cases still possess a cycle and anti-solar cases do not. Conversely, we have not found any possibility for sustaining a magnetic cycle with the traditional Babcock-Leighton flux-transport dynamos in the anti-solar differential rotation regime due to flux addition. Graphic interpretations are proposed in order to illustrate these cases. However, we find that hybrid models containing both prescriptions can still sustain local polarity reversals at some latitudes.}
   {We conclude that stars in the anti-solar differential rotation regime can sustain magnetic cycles only for very specific dynamo processes. The detection of a magnetic cycle for such a star would therefore be a particularly interesting constraint in  working to decipher what type of dynamo is actually at work in solar-type stars.}

    
    \titlerunning{Anti-solar differential rotation in mean-field solar-type dynamos}
    
   \maketitle
   
\section{Introduction}
The rotation of stars is a key ingredient in working to understand and characterize their dynamical nature. Indeed, the rotation and magnetism of stars are closely intertwined along their evolution, which directly impacts transport properties in their interior (\citealt{Emeriau-Viard&Brun2017}, \citealt{Brun+2021}).\\
Young solar-type stars begin their life by contracting and spinning up until they reach the zero-age main sequence (ZAMS). Then they enter a stage where they will spend most of their life-time, namely: the main sequence (MS). During this phase, their rotation $\Omega_*$ spins-down slowly, due to mass and angular momentum losses through the magnetized stellar wind (\citealt{Schatzman1962}, \citealt{WeberDavis1967}, \citealt{Mestel1968}). This wind braking (\citealt{Kawaler1988}, \citealt{Matt+2012}, \citeyear{Matt+2015}, \citealt{Reville+2015a}, \citealt{FinleyMatt2017}, \citealt{Vidotto2021} and references therein) causes the rotation of stars with similar masses to cluster around a similar evolution, which is particularly well described by Skumanich's law: $\Omega_*(t)\propto t^{-1/2}$ (\citealt{Skumanich1972}). It is therefore possible to determine stellar ages with measurements of their masses and rotation period, which is the basis of gyrochronology (\citealt{Barnes2003}, \citeyear{Barnes2007}). The magnetic activity of stars also follows a somewhat similar evolution. Fast and young rotating stars are found to possess a strong level of magnetic activity \citep{pizzolatoStellarActivityrotationRelationship2003}. The X-ray luminosity of stars is then found to decrease with their rotation period, that is, with their age (\citealt{wrightSTELLARACTIVITYROTATIONRELATIONSHIPEVOLUTION2011}, \citealt{reinersGENERALIZEDINVESTIGATIONROTATIONACTIVITY2014}). The same trends were more recently found in magnetic field estimates through Zeeman signatures by \cite{vidottoStellarMagnetismEmpirical2014} and \cite{seeNondipolarMagneticFields2019}. However, the magnetorotational evolution of stars older than the Sun is unclear at present. Indeed, some observers have found evolved solar-type stars following the Skumanich’s law (\citealt{LorenzoOliveira2018}, \citeyear{LorenzoOliveira2019}, \citealt{DoNascimento2020}), while others have not (\citealt{VanSaders2016}, \citeyear{VanSaders2019}, \citealt{Hall+2021}). It is plausible that a change in magnetic field sustained through dynamo processes could be at the source of these discrepancies \citep{metcalfeMagneticEvolutionDisappearance2017}, however, the magnetic field of slowly rotating stars is particularly challenging to detect. As a result, we propose a theoretical exploration of the possible dynamo states for slow rotators based on mean-field models.\\
Besides its temporal evolution, the star rotation profile can evolve in terms of latitude. It can be close to uniform in the meridional plane or differentially distributed in the convection zone (\citealt{Brun+2021}). Differential rotation (DR) is indeed one of the most important ingredients for dynamo action in stellar interiors, spanning the convective envelope to such an extent that the equator rotates faster than the poles for the Sun \citep{Thompson2003}. Following the pioneering work of \citet{Gilman1977} and \citet{GilmanGlatzmaier1981}, several authors have recently highlighted that the type of DR profile realized in a stellar turbulent rotating convective envelope is directly linked with its effective Rossby number (\citealt{Matt+2011}, \citealt{Guerrero+2013}, \citealt{Gastine+2014}, \citealt{Kapyla+2014}, \citealt{Simitev+2015}, \citealt{Brun+2017}, \citealt{Karak+2018}, \citealt{Brun+2021}). The dimensionless fluid Rossby number, $Ro_{\rm f}=\omega/2\Omega_*$, with $\omega$ as the vorticity,  quantifies the effect of rotation on turbulent convection. The aforementioned studies (see \citealt{BrunBrowning2017} for a review)  consistently find a qualitative change of the differential rotation profile between intermediate and slow rotators. Intermediate rotators are found to possess a solar-like differential rotation (fast equator, slow poles) and slow rotators to possess an anti-solar differential rotation (slow equator, fast poles). In \cite{Brun+2021}, this transition is found around $Ro_{\rm f}$=1 (in this case the solar $Ro_{\rm f}$=0.9).\\
As yet, there has not been any statistically strong detection of anti-solar DR for cool MS-stars. \cite{Benomar+2018} analyzed the DR of 40 solar-like stars using asteroseismology and only found some likely anti-solar targets so far. \cite{Strassmeier+2003}, \cite{Weber+2005} and \cite{Kovari+2007b} reported the detection of anti-solar DR thanks to Doppler imaging spectroscopy. These detections were nevertheless made for evolved stars, such as K giants, and more recently for subgiants \cite{harutyunyanAntisolarDifferentialRotation2016}. Finally, \cite{reinersDifferentialRotationStars2007} attempted the detection of slowly rotating M-dwarfs and could not conclude on whether its targets possessed a solar or anti-solar DR. The detection of anti-solar DR in MS cool stars is therefore still pending and will require the scrutiny of the most promising targets we know (Noraz et al. in prep). It can nevertheless be expected that a solar-type star might transit toward an anti-solar DR state if it becomes old enough. This transition could affect stellar activity and may stand as the origin of the enhanced stellar activities, which were recently observed by \cite{brandenburgEnhancedStellarActivity2018} in the high Rossby regime, where the observation of an anti-solar DR profile is anticipated.\\
Several studies of global 3D MHD simulations have discussed non-linear dynamos realized under anti-solar DR states. \citet{Karak+2015} first found irregular and faint cycles with polarity changes for only few latitudes, when \citet{Varela+2016} did not. Clearer and global reversals are however present for anti-solar rotation in \citet{Viviani+2018} and (\citeyear{Viviani+2019}) simulations, but mostly irregular and only near the DR transition regime. On the other hand \citet{Warnecke+2018} and \citet{Strugarek+2018} reported mostly stationary dynamos for this particular rotation state. This is also the case for \cite{Brun+2021}, extending the study initiated by \citet{Varela+2016}.\\
In this context, we propose a numerical experiment to study the impact of an anti-solar rotation regime on the dynamo process and, in particular, the presence of magnetic cycles or otherwise. Global 3D MHD simulations are useful because they self-consistently treat the flows and dynamo magnetic fields. However, they are numerically expensive. In  order  to  explore  a  broad  parameter space, we propose 2D dynamo models using the mean-field theory presented in Section 2.1. Such models have been explored by \cite{Dube&Charbonneau2013ApJ} using mean-field coefficients extracted from 3D MHD prescriptions. Likewise, \cite{karak_stellar_2020} have considered anti-solar models and found they could trigger magnetic cycles, but only with strong $\alpha$-effect coupled with $\Omega$-quenching. These authors obtained reversals in a $\alpha^2\Omega$ dynamo and argued that this could not be the case with classical kinematic $\alpha\Omega$ models when considering a positive $\alpha$-effect in the northern hemisphere. Here, we actually chose to consider standard $\alpha\Omega$ kinematic models and we will complete these previous studies with the following layout.\vspace{\baselineskip}\\
We first present the chosen models along with their physical ingredients in Section \ref{sec:MF-Model}, in particular how the anti-solar DR profile is chosen. Then in Section \ref{sec:DR_influence}, we construct two solar DR reference dynamos, using standard prescriptions, namely $\alpha\Omega$ (\citealt{Parker1955}) and Babcock-Leighton (BL) mechanisms (\citealt{Babcock1961}, \citealt{Leighton1969}) in an interface dynamo approach \citep{Parker1993}. We then present the same models with an anti-solar DR. Next, in Section \ref{sec:RobResults}, we study the role of the $\alpha$-term location and the meridional circulation (MC) on our results. We finally discuss the context and limitations of our choices in Section \ref{sec:Context} and present our conclusions in Section \ref{sec:ccl}.
\section{Stellar mean-field dynamo model}\label{sec:MF-Model}
In this study, we chose to use a mean-field approach to model stellar dynamos (\citealt{Roberts1972}, \citealt{KrauseRaedler1980}). We refer interested readers to the review in \cite{Charbonneau2020LRSP} on solar and stellar dynamos using this approach. We first present the model setup, then we describe the numerical implementation that is used, and, finally, we detail physical ingredients prescriptions chosen for this study. 

\subsection{Mean-field equations}\label{sec:MF-Equations}
We first take the induction equation:
\begin{eqnarray}
    \dfrac{\partial\bm{B}}{\partial t}=\nabla\times (\bm{v} \times\bm{B})-\nabla\times(\eta_m\nabla\times\bm{B})\;.
    \label{eq:induction}
\end{eqnarray}
This equation governs the evolution of the magnetic field $\bm{B}$ in response to a flow field $\bm{v}$, acting against the ohmic dissipation characterized by the microscopic magnetic diffusivity $\eta_m$.\\
The well-known development of the mean-field dynamo model can be found in \citet{moffatt_magnetic_1978} or \citet{Charbonneau2020LRSP}. Here, we briefly report the different steps as follows: We first decompose both the velocity and the magnetic fields into their mean ($\langle\bm{v}\rangle$, $\langle\bm{B}\rangle$) and fluctuating ($\bm{b'}$, $\bm{v'}$) components. Our study then focuses on the mean magnetic part $\langle\bm{B}\rangle$ evolution, corresponding to the large scale magnetic field. This large-scale component is impacted by small-scale turbulence through the electromotive force (EMF) term $\epsilon=\langle\bm{u'}\times\bm{b'}\rangle$. This term will here be parameterized in a kinematic approach with two scalar terms $\alpha$ and $\beta$ (respectively creating and destructing large-scale $\langle\bm{B}\rangle$), resulting from an asymptotic expansion of $\epsilon$ in the so-called first order smoothing approximation (FOSA, \citealt{BrandenburgSubramanian2005}). We then note the effective magnetic resistivity $\eta=\eta_m+\beta$, and refer to the mean magnetic field, $\bm{B}$, and mean velocity field, $\bm{v}$, by omitting the angle brackets $\langle$ $\rangle$ for clarity. As we are working in spherical coordinates ($r,\theta,\phi$), under the assumption of axisymmetry, a convenient poloidal and toroidal decomposition of the mean fields is:
\begin{eqnarray}
    \bm{B}(r,\theta,t)&=&\nabla\times (A_\phi(r,\theta,t) \hat{\bm{e}}_\phi)+B_\phi(r,\theta,t) \hat {\bm{e}}_\phi, \\
\bm{v}(r,\theta)&=&{\bm{u}_p}(r,\theta) + r\sin\theta \, \Omega(r,\theta) \hat {\bm{e}}_\phi.
\end{eqnarray}
As our approach is kinematic, we do not assume any fluctuations in time of the differential rotation $\Omega$ and of the meridional circulation $\bm{u}_p$. Reintroducing this decomposition into (\ref{eq:induction}), we get the two coupled partial differential equations:
\begin{eqnarray}
    \label{eq:Ap}
    \dfrac{\partial A_\phi}{\partial t}=\dfrac{\eta}{\eta_{t}} \left(\nabla^2-\dfrac{1}{\varpi^2}\right) A_\phi -R_{\rm m}\dfrac{\bm{u}_p}{\varpi}\cdot\nabla(\varpi A_\phi) +C_\alpha\alpha B_\phi+C_sS,
\end{eqnarray}
\begin{eqnarray}
    \label{eq:Bp}
    \dfrac{\partial B_\phi}{\partial t}&=&\dfrac{\eta}{\eta_t} (\nabla^2-\dfrac{1}{\varpi^2})B_\phi \nonumber - R_{\rm m}\varpi \bm{u}_p\cdot\nabla(\dfrac{B_\phi}{\varpi})-R_{\rm m}B_\phi\nabla\cdot\bm{u}_p \nonumber \\
    &+&C_\Omega\varpi(\nabla\times(A_\phi\hat{\bm{e}}_\phi))\cdot\nabla\Omega\;,
    +\frac{1}{\varpi}\dfrac{\partial(\varpi B_\phi)}{\partial r}\dfrac{\partial(\eta/\eta_{t})}{\partial r},
\end{eqnarray}
where the cylindrical radius $\varpi=r\sin\theta$ and $\eta_t$ is the turbulent magnetic diffusivity (constant value taken by $\eta(r)$ in the convection zone). For Babcock-Leighton models, a non-local surface term $S$ is added to mimic the rise of buoyant magnetic structures in the convection zone (\citealt{Babcock1961}, \citealt{Leighton1969}, \citealt{choudhuriSchusslerDikpati1995}, \citealt{Charbonneau2020LRSP} and references therein).\\
These equations are written in a dimensionless form by choosing: the solar radius, $R_\odot,$ as the length scale and the diffusion time, $R_\odot^2/\eta_t$, as the timescale. We recall here that the induction equation in the kinematic regime leads to an arbitrary vector potential, $A_\phi,$ and magnetic field, $B_\phi$.\\
This, in turn, leads to three dimensionless numbers, which will serve as our control parameters: $C_\Omega=\Omega_{Eq}R_\odot^2/\eta_t$, $C_\alpha=\alpha_0R_\odot/\eta_t$, and $R_{\rm m}=u_0R_\odot/\eta_t$. Here, $\Omega_{Eq}$ is the amplitude of rotation at the equator; $\alpha_0$ and $u_0$ are the maximum amplitudes of the $\alpha$-effect term and the meridional flow, respectively. For Babcock-Leighton (BL) models, the added  $S$ term leads to a fourth dimensionless parameter: $C_s=s_0R_\odot/\eta_t$, where $s_0$ is the amplitude of the BL effect.\\
In this study, we chose to focus on $\alpha\Omega$ or Babcock-Leighton (BL) dynamos. Hence, an $\alpha$-effect will only be considered for the generation of the poloidal field and not for the toroidal field. We discuss this choice in Section \ref{sec:Context}.

\subsection{Numerical domain and boundary conditions}\label{sec:Numerical}
We used the same boundary conditions for all the simulations described in this paper. Assuming axisymmetry, Equations \ref{eq:Ap} and \ref{eq:Bp} are solved in the meridional plane with the colatitude $\theta\in[0;\pi]$ and the normalized radius $r\in[0.6;1]$. At $\theta=0$ and $\theta=\pi,$ we imposed a regularity by setting homogeneous conditions such that $A_\phi$ and $B_\phi$ are null. A perfect conductor condition is considered at $r_{b}=0.6$, which gives:
$$A_\phi=0 \; \text{and} \; \dfrac{\partial(rB_\phi)}{\partial r}=0.$$
Finally, the upper boundary condition at $r_{t}=1$ smoothly matches a potential field solution as the region $r>1$ represents a vacuum:
$$\left(\nabla^2-\dfrac{1}{r^2\sin^2\theta}\right)A_\phi=0 \; \text{and} \; B_\phi=0.$$
The magnetic field is initialized with a weak dipole and, thus, $B_\phi (t=0)=0$. We used the STELEM code (Emonet \& Charbonneau (1998), private communication) to solve the mean-field equations, by employing a first-order finite elements method in space, along with a third order scheme in time (see Appendix A of \citet{jouve_role_2007} for more details). All simulations proposed in this paper were performed with a $256\times128$ resolution (latitudinally vs. radially respectively) in order to reach a good numerical convergence. The accuracy of the dynamo solution is here mainly constrained by the latitudinal number of grid points.

\subsection{Physical ingredients}\label{PhysIngred}
In this study, we use relatively simple dynamo ingredients that have been found to reproduce the main characteristics of the large-scale solar dynamo. They are described in the following subsections.

\subsubsection{Toroidal magnetic field generation:  The $\Omega$-effect}\label{sec:OmegaEffect}
The form of $\Omega(r,\theta)$ was chosen to mimic the solar differential rotation profile, as inferred from helioseismology \citep{Thompson2003}. The analytic fit used here is inspired from \citet{DikpatiCharbonneau1999}, where the radial shear is maximal in the tachocline:
\begin{eqnarray}
    \label{eq:Thomson_DR}
    \Omega(r,\theta) = \dfrac{\Omega_C}{\Omega_{Eq}}&+&\dfrac{1}{2}\left[ 1 + {\rm erf} \left(2\dfrac{r-r_{\rm BCZ}}{d}\right)\right]\\
    &\times&\left(1 + \dfrac{a_2}{\Omega_{Eq}}\cos^2(\theta)+\dfrac{a_4}{\Omega_{Eq}}\cos^4(\theta)-\dfrac{\Omega_C}{\Omega_{Eq}}\right),\nonumber
\end{eqnarray}
with $\Omega_C$ the core angular velocity. The tachocline is placed at $0.7 R_\odot$ and its thickness, $d,$ is expected to be $0.025R_\odot$. Symmetry considerations allow us to create a typical anti-solar differential rotation profile by modifying the coefficients of Equation \ref{eq:Thomson_DR}.
\begin{table}
\centering
\caption{Ratios between coefficients of Equation \ref{eq:Thomson_DR} for solar and anti-solar differential rotation profiles.}
\begin{tabular}{|c|c|c|}
    \hline
     & $\odot$ & anti-$\odot$\\
    \hline
    $\Omega_C/\Omega_{Eq}$ & 0.93944 & 1.06056\\
    $a2/\Omega_{Eq}$ & -0.136076 & 0.136076\\
    $a4/\Omega_{Eq}$ & -0.145713 & 0.145713\\
    \hline
\end{tabular}
\label{tab:DR_coefs}
\end{table}\\
\begin{figure*}
        \centering
        \includegraphics[width=\linewidth]{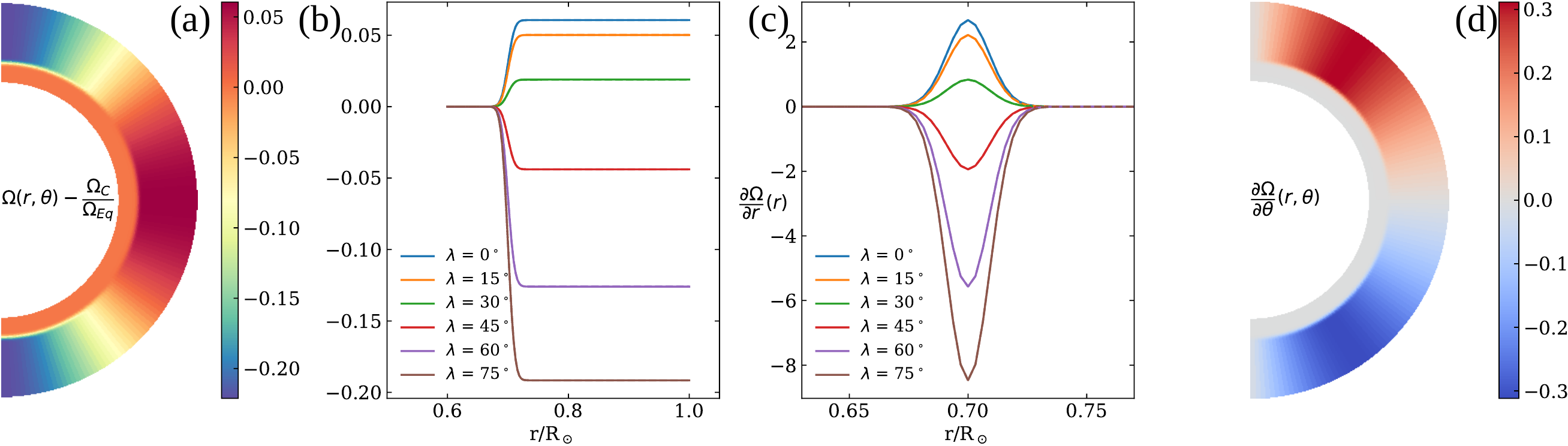}
        \caption{Solar differential rotation shown with a meridional cut $(a)$ and with radial profiles at different latitudes $\lambda$ (b); $(c)$ and $(d)$ show the radial and latitudinal shear respectively, responsible for the so-called $\Omega$-effect. Profiles are represented in the frame rotating to $\Omega_C$.}
        \label{fig:SolDR}
\end{figure*}
\begin{figure*}
    \centering
    \includegraphics[width=\linewidth]{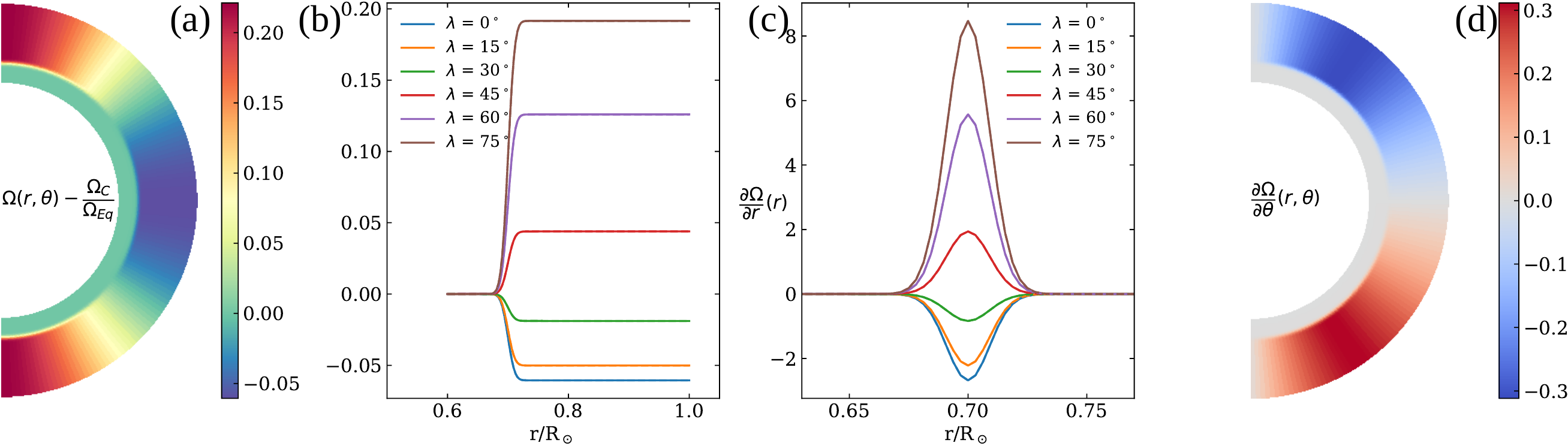}
    \caption{Anti-solar differential rotation, responsible for the so-called $\Omega$-effect. Same details as in Figure \ref{fig:SolDR}.}
    \label{fig:AntisolDR}
\end{figure*}
We show and detail these two profiles in panels (a) in Figures \ref{fig:SolDR} and \ref{fig:AntisolDR}, where they are represented in the core frame rotating to $\Omega_{C}$. Panels (b) represent their radial profiles taken at different latitudes, which have been obtained through Equation \ref{eq:Thomson_DR}, using the coefficients in Table \ref{tab:DR_coefs}. First, we have in both cases a solid body rotation for $0.6<r<0.7$, where we find the stable radiative zone. This lower region is surrounded by the convective zone (CZ) where the rotation is differential, that is, the angular velocity varies along latitudes, until the surface $r=1$ (we recall here that in 2D mean-field dynamos convective motions are not resolved). For the solar (resp. anti-solar) case, the equatorial part of this upper region rotates faster (resp., slower) than the core, contrary to the poles, which rotate slower (resp., faster). In both solar and anti-solar cases, we note a co-rotation latitude around $\lambda\sim35^\circ$, namely, a latitude at which both the upper and lower regions rotate at the same rate. Although we have a latitudinal shear $\partial_\theta\Omega$ in most of the convection zone, the radial shear $\partial_r\Omega$ is confined around $r=0.7$ in an area called the tachocline. We finally note a stronger shear $\nabla\Omega$ at higher latitude, which is chosen here to follow the DR profile inferred from helioseismology. We can see it in panels (c) and (d), representing the radial shear and the latitudinal shear, respectively. The former is represented with radial profiles taken at different latitudes.\\
Helioseismic inversions from \citet{Thompson2003} show that the Sun’s radiative zone has a solid body rotation, just below the tachocline. A convective zone is found to surround them until the stellar surface, where the rotation becomes differential. Indeed, non-linear convective motions redistribute the angular momentum (\citealt{Brun2002}, \citeyear{Brun+2017}). In the solar case, this redistribution transports angular momentum toward the equator. It leads to different surface latitudes spinning at different velocities, from a fast equator to slow poles, including a latitude in co-rotation (i.e.,  same angular velocity) with the radiative core. Moreover, these same simulations also show that anti-solar profiles may exist. In this case, the radiative interior in still in solid-body rotation, but the differential rotation of the convective zone is latitudinally reversed from the solar regime. Thus, the equatorial convective region is found this time to rotate slower than the core, which is contrary to polar convective regions, which spin faster.\\
The coefficients in Table \ref{tab:DR_coefs} were therefore chosen to reproduce these profiles, while keeping absolute shear amplitudes equal in the last term of Equation \ref{eq:Bp}, namely, the $\Omega$-effect. The absolute differences have thus been conserved in both profiles, between $\Omega_{Eq}$ and $\Omega_C$ controlling the radial shear, along with the absolute amplitude of $a2/\Omega_{Eq}$ and $a4/\Omega_{Eq}$ controlling the latitudinal shear. Thus, $|\nabla\Omega(r,\theta)|$ is fully conserved, as we can see in Figures \ref{fig:SolDR} and \ref{fig:AntisolDR}.\\
We note that this anti-solar profile differs from the one used in \citet{karak_stellar_2020}. In this similar study using an $\alpha^2\Omega$ dynamo model, the value of the anti-solar coefficient $\Omega_C/\Omega_{Eq}$ is kept at 0.93944 (as in the solar DR profile). It results in a radiative zone rotating more slowly than the whole convective zone, with no co-rotation latitude in the latter, which seems to be in contradiction with anti-solar differential rotation profiles found in recent 3D self-consistent simulations (\citealt{Dube&Charbonneau2013ApJ}, \citealt{Brun+2017}, 2021). As $\Omega$ gradients play a major role in the $\Omega$-effect, this could seriously impact results of such mean-field models.

\subsubsection{Poloidal magnetic field generation}\label{sec:PolGene}
\begin{figure}
    \centering
    \includegraphics[width=\linewidth]{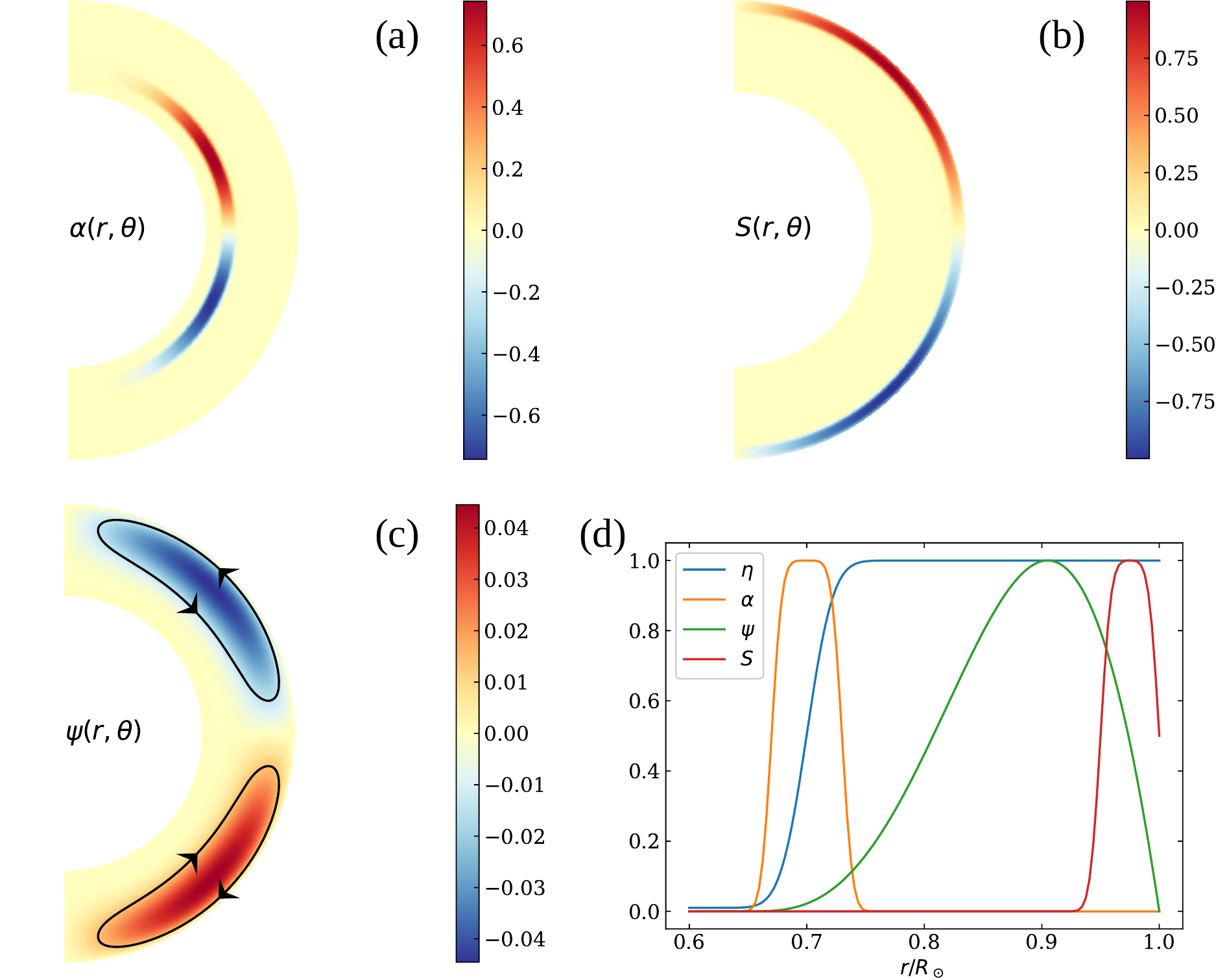}
    \caption{Physical ingredients: Meridional profiles of (a) $\alpha$-effect, (b) BL-effect, as well as (c) the stream function $\psi$ (MC) and (d) normalized radial profiles of the latter quantities, taken at $\theta=\pi/4$, along with the magnetic diffusivity $\eta$. These profiles result from Equations \ref{eq:alpha_profile_surf}, \ref{eq:BL_profile_surf}, \ref{eq:stream_function}, and \ref{eq:eta}, respectively. The $\alpha$-term is here confined between $(r_1,r_2)=(0.67,0.73)$, as in the reference cases. The BL source term $S$ is shown for the case of $B_\phi(r_{\rm BCZ},\theta,t)=1$. The sign of the meridional circulation, $\bm{u}_p$, is represented with arrows on a black iso-contour of $\psi$.}
    \label{fig:phys_ingred}
\end{figure}

Here, we consider two types of mean-field dynamo: $\alpha\Omega$ and Babcock-Leighton (BL), which differ by the poloidal source term. For our $\alpha\Omega$ dynamos, the form of the $\alpha$-term profile has been inspired from \citet{Bushby2006} and is written as: 
\begin{align}
    \alpha(r,\theta,t)=&\;\dfrac{3\sqrt{3}}{8}\left[1+\text{erf}\left(\dfrac{r-r_1}{0.01}\right)\right]\nonumber\left[1-\text{erf}\left(\dfrac{r-r_2}{0.01}\right)\right]\\
    &\times\left[1+\left(\dfrac{B_\phi (r,\theta,t)}{B_0}\right)^2\right]^{-1}\cos{\theta}\sin^4\theta.
    \label{eq:alpha_profile_surf}
\end{align}
This profile is represented in Figure \ref{fig:phys_ingred} (panel (a)) and allows us to take an $\alpha$-effect that can be localized radially within $[r_1,r_2]$. This choice of location will be discussed more specifically in Sections \ref{sec:DR_influence} and \ref{sec:RobResults}. However, the $\alpha$-effect parametrizes the cyclonic turbulence, so we always assume that its latitudinal variation is constrained by its dependence upon the Coriolis force, which makes it antisymmetric with respect to the equator.\\
As we are in a kinematic regime, there is no retro-action of the magnetic field on the velocity. Thus, a quenching term is introduced to prevent the magnetic energy from growing endlessly, with the parameter $B_0$ arbitrarily set to $10^4$.
\\
\indent In Babcock-Leighton dynamo models, the poloidal field is generated by the inclination of magnetic loops emerging at the solar surface; thus, the source has to be confined to a thin layer just below the surface. Since the rising-time of magnetic loops through the convection zone is short compared to the evolution of the large-scale $\bm{B}$, the process is fundamentally non-local. Hence, the surface source term depends on the variation of $B_\phi$ at the base of the convection zone (BCZ). Here, we set its position to $r_{\rm BCZ}=0.7$. As in the $\alpha$-effect, a quenching term is introduced here and we Thus, we use the following formulation:\\
\begin{align}
    \label{eq:BL_profile_surf}
    S(r,\theta,t)=&\;\dfrac{1}{2}\left[1+\text{erf}\left(\dfrac{r-0.95}{0.01}\right)\right]\left[1-\text{erf}\dfrac{r-1}{0.01}\right]\\
    &\times\left[1+\left(\dfrac{B_\phi (r_{\rm BCZ},\theta,t)}{B_0}\right)^2\right]^{-1}\cos\theta\sin\theta \;B_\phi(r_{\rm BCZ},\theta,t).\nonumber
\end{align}
\indent In BL flux-transport dynamo models, meridional circulation (MC) can be used to  dynamically link the two sources of the magnetic field, which are the $\Omega$-effect at the base of the convection zone (discussed in the previous subsection) and the BL-term at the surface. Its profile will be constructed as in \citet{jouve_solar_2008}: a large single cell per hemisphere, directed poleward at the surface and equatorward deep in the CZ. It will penetrate a little below the tachocline and vanish at $r_o=0.6$. Thus, the stream function $\psi$ will be taken as:
\begin{equation}
    \psi(r,\theta)=-\dfrac{2}{\pi}\dfrac{(r-r_o)^2}{(1-r_o)}\sin\left(\pi\dfrac{r-r_o}{1-r_o}\right)\cos{\theta}\sin{\theta},
    \label{eq:stream_function}
\end{equation}
following the relation $\bm{u}_p=\nabla\times(\psi\bm{e}_\phi)$. As illustrated in panel (c) of Figure \ref{fig:phys_ingred}, positive (resp., negative) values of $\psi$ means clockwise (resp., counter-clockwise) $\bm{u}_p$ circulation. This can be recovered through a right-hand rule, with respect to the spherical basis vector $\bm{e}_\phi$. This stream function $\psi$ leads to a bottom-top CZ $u_{\theta,bot}/u_{\theta,top}=0.352$ contrast, normalized at $u_\theta(r=1,\theta=45^\circ)=1$.

\subsubsection{Ohmic diffusion}
Finally, we  assume that the diffusivity in the convective zone $\eta_t$ is dominated by its turbulent contribution, whereas in the radiative interior: $\eta_C << \eta_t$. We take $\eta_C=0.01\eta_t$ with $\eta_t=10^{11}$ cm$^2$.s$^{-1}$ for the models presented in this paper. We chose this value of $\eta_t$ such that our reference BL model produces a magnetic cycle of 22 years when $R_m=500$, which corresponds to a MC amplitude of $u_0=7.2$ m.s$^-1$. For the sake of consistency, we kept the same value of $\eta_t$ for all the other models.\\
The convection zone is distributed from the surface $r=1$ to its base $r_{\rm BCZ}=0.7$. As shown in panel (d) of Figure \ref{fig:phys_ingred}, we used an error function in order to smoothly match the two different values at this radius such that\begin{equation}
    \Tilde{\eta}(r)=\dfrac{\eta}{\eta_t}=\dfrac{\eta_C}{\eta_t}+\dfrac{1}{2}\left(1-\dfrac{\eta_C}{\eta_t}\right)\left[1+\text{erf}\left(2\dfrac{r-r_{\rm BCZ}}{0.025}\right)\right]\;.
    \label{eq:eta}
\end{equation}

\section{Dynamo states for solar and anti-solar DR}
\label{sec:DR_influence}
We recall here that we consider two types of dynamo models in this study. For readers seeking context, we suggest comparing them with representative models of \citeauthor{Charbonneau2020LRSP} (\citeyear{Charbonneau2020LRSP}, Sections 4.2.10 and 5.4.2).\vspace{1\baselineskip}\\
In this section, we assume the $\alpha$-effect to be driven by some tachocline-based instability, as in \cite{Bushby2006}. Different types of such instabilities are discussed in \citeauthor{Charbonneau2020LRSP} (\citeyear{Charbonneau2020LRSP}, Section 4.5). Hence, the $\alpha$-term will be confined within the tachocline between $r_1=0.67$ and $r_2=0.73$ (see Figure \ref{fig:phys_ingred}, panel \textit{b}).\\
We note here that the sign of the parametrized $\alpha$-effect can be related in the kinematic regime to the averaged kinetic helicity and the convective turnover timescale (see \citealt{BrandenburgSubramanian2005}):
\begin{equation}
    \alpha\sim-\dfrac{\tau_C}{3}<\bf{u'}\cdot\mbox{\boldmath $\nabla$}\times \bf{u'}>
    \label{eq:alpha_kin}
,\end{equation}
Vertical motions of the convection zone (CZ) are cyclonic (counterclockwise in the northern hemisphere) and change sign as they reach the bottom of this zone (e.g., \citealt{Miesch2000}, \citealt{Duarte2016MNRAS}, \citealt{Charbonneau2020LRSP}). This leads to a change of sign of $\alpha$-effect when crossing the bottom of the CZ; Thus, $C_\alpha$ will be negative for $\alpha$-profiles localized at the tachocline -- and positive otherwise (see Table \ref{tab:transBush}).\\

\subsection{Solar dynamo models}\label{sec:SuRef}
Our solar reference cases have to model observational tendencies of the Sun, illustrated with so-called "butterfly diagram" (magnetogram evolution vs. latitude) as in Figure 17 of \citet{hathawaySolarCycle2015}. First, an 11-year activity cycle due to the emergence of magnetic structures, which are typically dipolar. Throughout the cycle, the emergence of these structures will migrate towards the equator. At the same time, the trailing polarity will diffuse towards the pole, giving rise to a polar branch. This branch will then progressively cancel the magnetic flux at the pole, resulting from the previous activity phase. The global dipole of the solar magnetic field will then reverse, as will the polarization of emerging magnetic structures in the next phase of activity. This leads to a 22-year magnetic cycle.
\subsubsection{$\alpha\Omega$ solar reference model}
\begin{figure}
    \centering
    \includegraphics[width=\linewidth]{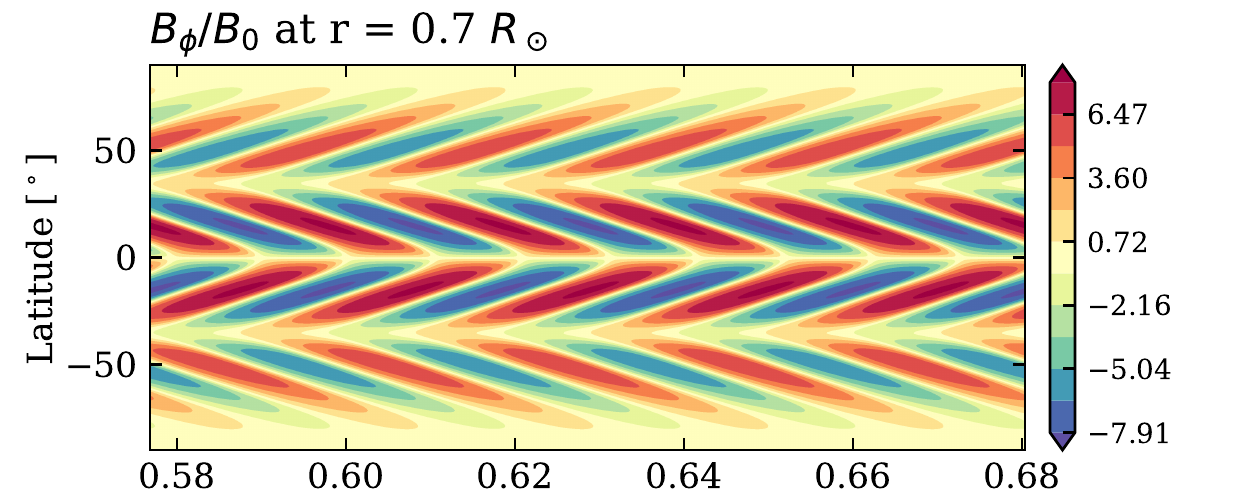}
    \includegraphics[width=\linewidth]{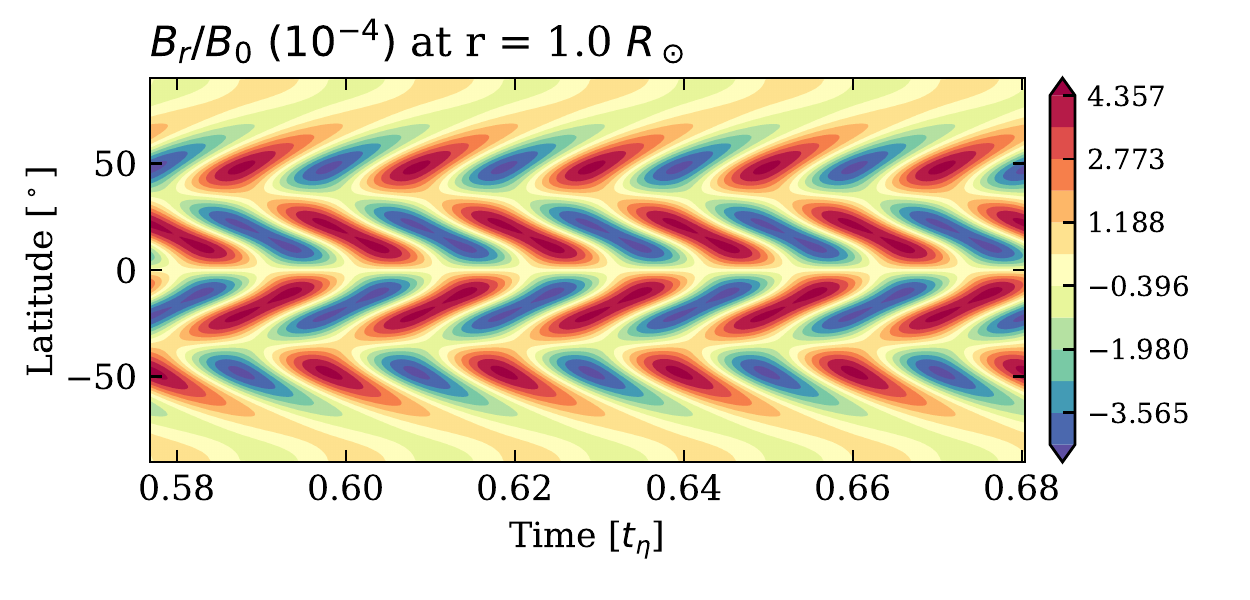}
    \caption{Butterfly diagram of Solar $\alpha\Omega$ model with $C_\alpha=-15$ and $D=-2.1\times10^6$, for $B_\phi$ at $r=0.7$ and $B_r$ at the surface. We recall that for the purposes of this study, 22 years correspond to 0.014 $t_\eta$.}
    \label{fig:Sun_AO}
\end{figure}

Assuming a solar differential rotation (fast equator, slow poles), we first computed an $\alpha\Omega$-dynamo model with $C_\Omega=1.4\times10^5$ and $C_\alpha=-15$. These values have been chosen to reproduce cycle periods of a few tens of years with an equatorial rotation rate of $\Omega_{Eq}/2\pi = 456$ nHz for $\eta_t = 10^{11}$cm$^2$.s$^{-1}$.
Butterfly diagrams are shown in Figure \ref{fig:Sun_AO} for $B_\phi$ and $B_r$ (respectively) at the base and the surface of the convection zone. Both of them show a dipolar configuration over the meridional plane. The bottom panel representing $B_r$ first shows that this dynamo model reproduces well the observed surface dynamics of the Sun. The cyclic equator-ward migration of the emergence of magnetic patterns along with their trans-equatorial cancellation with the opposite polarity arriving from the opposite hemisphere is well modeled. The same holds true for the polar branch leading to a cyclic global polarity reversal, with a period of 32 years in this case. We did not seek to fine-tune the model to set $P_{\rm cyc}=22$ years (see Figure \ref{fig:SunModelAOtuned} in the appendix for such a case) as the exact value of $P_{\rm cyc}$ has no impact on our results. This is discussed in more detail in Section \ref{sec:AstroCont}.\\
We go on to note in the top panel of Figure \ref{fig:Sun_AO} that the generation of the toroidal magnetic field $B_\phi$ at the tachocline has two branches in each hemisphere.
They propagate away from each other by sharing the same origin, situated at the co-rotation latitude. Moreover, the prescribed radial shear $\partial_r\Omega$ gives them similar propagation speed (see Parker-Yoshimura sign rule in Equation \ref{eq:P-Y prop}). Thus, it  leads to them being phase-locked. We may note that both branches have here the same polarities in one hemisphere. As they are inverted with respect to the equator, a dipolar magnetic topology is thus found to dominate throughout the meridional plane, which is true for all magnetic field components (see Figure \ref{fig:MeridCuts} in the appendix for details).\\
The equatorial branch is slightly stronger than the polar one. Indeed, the form of the $\alpha$-effect is concentrated toward the equator by the $\sin^4\theta$ factor (Equation \ref{eq:alpha_profile_surf}), generating more $B_{pol}$ there, which is subsequently available for toroidal generation. Nonetheless, the polar branch still remains prominent in this model as its propagation originates from a position ($r,\theta$), where the $\alpha$-effect is still strong.\\
\subsubsection{Babcock-Leighton flux-transport solar reference model}
\begin{figure}
    \centering
    \includegraphics[width=\linewidth]{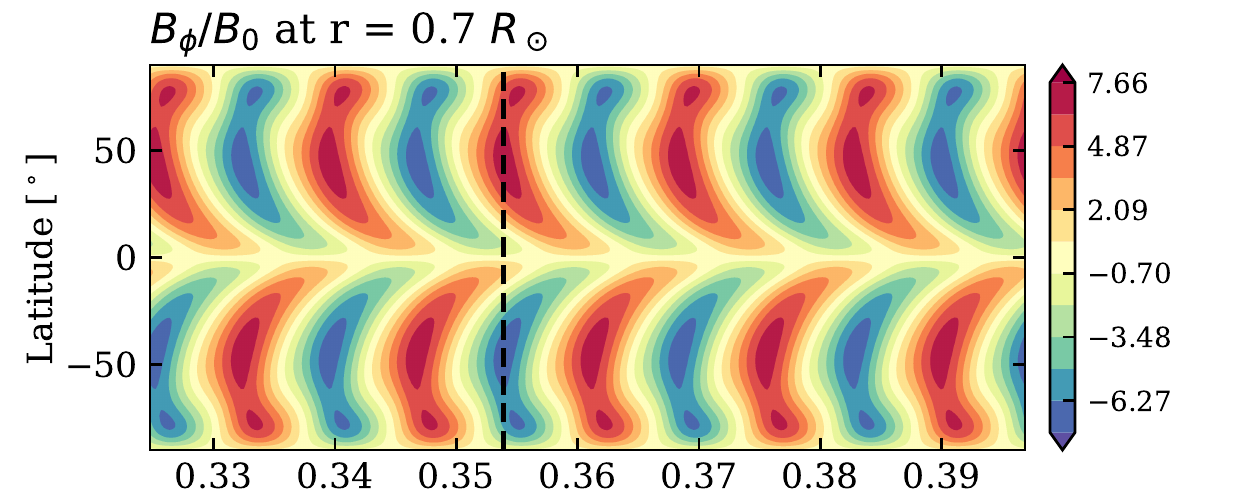}
    \includegraphics[width=\linewidth]{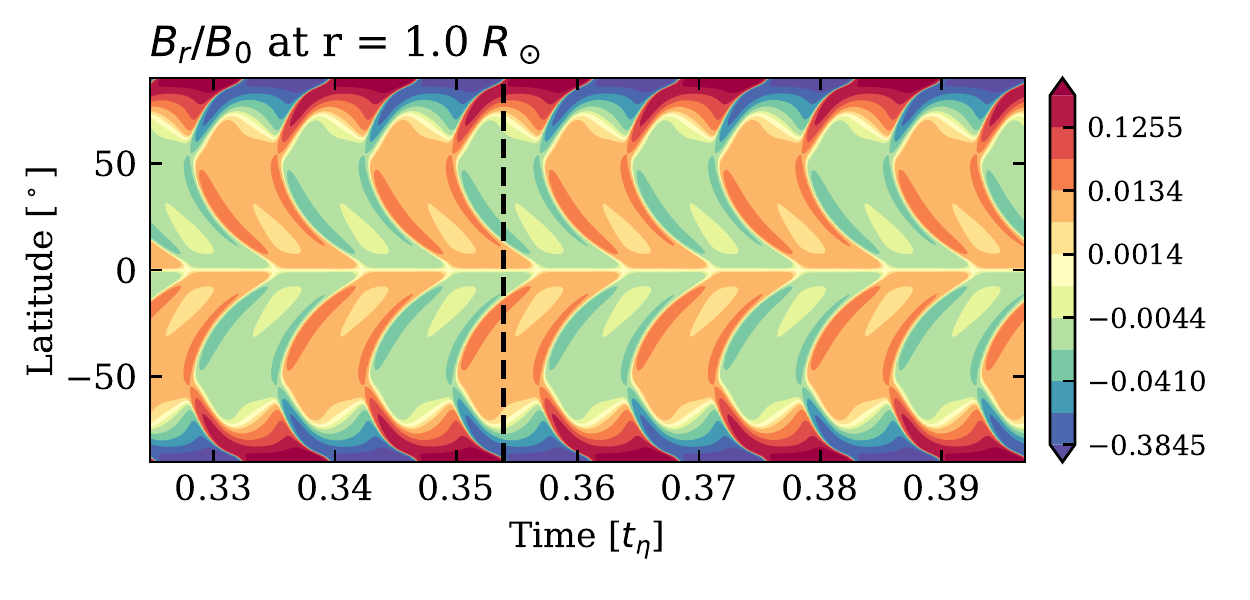}
    \caption{Butterfly diagram of solar Babcock-Leighton model with $C_S=30$ and $D=4.2\times10^6$, for $B_\phi$ at $r=0.7$ and $B_r$ at the surface. As the radial component diagram is much more contrasted here, we propose some contours, using consistent colors, to illustrate weak patterns that are not saturated enough. The vertical dashed line corresponds to the reversal of poloidal field from negative to positive at the poles, coinciding with the positive maximum of toroidal field at the base of CZ. A logarithmic color scale has been chosen to represent the bottom panel.}
    \label{fig:Sun_BL}
\end{figure}

\indent We then constructed a BL flux-transport model that also reproduces a Sun-like dynamo. We used the same DR profile and parameters. The meridional flow and the BL term were chosen  as $R_{\rm m}=500$ and $C_s=30$. As with the previous $\alpha\Omega$ model, this BL dynamo reproduces well several aspects of the solar cycle, notably its period of 22 years (by considering once again $\Omega_{Eq}/2\pi = 456$ nHz and $\eta_t = 10^{11}$cm$^2$.s$^{-1}$). Indeed, we see at the surface a dipolar magnetic structure on the lower panel of Figure \ref{fig:Sun_BL}, with two $B_r$-branches in each hemisphere, with analogous dynamics toward the equator and the pole. The same topology is shown for $B_\phi$ on the top panel, again showing  two branches going apart from one another with the same polarities in each hemisphere. We next note a phase shift of $\pi/2$ between the deep toroidal field and the surface polar field. The vertical black line on the butterfly diagrams in Figure \ref{fig:Sun_BL} show a polarity change of the polar field from negative to positive, when a positive value for the toroidal field is maximal in the equatorial branch at the base of the convection zone.\\
Additionally, this dynamo mechanism reveals its own particularities. First, the poloidal field is now generated by the BL-source term close to the surface of the star. Thus, more of the poloidal field will be available for toroidal generation through the latitudinal shear near the surface. Second, the equatorial branches at the base of the CZ are more extended than the polar ones. This is the signature of the meridional circulation dragging the toroidal field equatorward at the base of the convection zone. Hence, we clearly see that the advection of the magnetic field is dominating its diffusion in the domain. The signature of the MC is also seen on the bottom panel with $B_r$ at the surface, where the field advection is now poleward. Indeed, the polarity inversion line (here, the latitude at which branches separate) of each hemisphere is at a higher latitude than what is expected from the $\alpha\Omega$ model. We illustrate in Figure \ref{fig:MeridCuts} this dynamics with time series of meridional cuts of magnetic components. Moreover, we also note  stronger values of the surface radial magnetic component at both poles. Indeed, the meridional circulation accumulates magnetic field over there, while the diffusion is too weak to avoid it. When better agreements with solar observations are needed, diffusion can be enhanced at the surface (see \citealt{dikpatiFluxTransportSolar2002} and \citealt{hottaIMPORTANCESURFACETURBULENT2010} for instance). For the sake of simplicity, such a refinement is not needed for our purpose of studying anti-solar DR dynamo and seemed not to be relevant to the existence or non-existence of cycles. However, this aspect could be explored in future works.\\

\subsection{Dynamos with anti-solar DR}\label{sec:AnRef}
Now that our physical ingredients are set up to reproduce what we currently know about Sun-like dynamos, we take our anti-solar rotation profile and apply it to these two dynamo models. We keep here the same $C_\alpha$ value without changing its sign. Indeed, recent 3D non-linear simulations show that the behavior of helicity is globally identical for solar and anti-solar rotation profiles, and that the convective turnover time tends to keep the same order of magnitude between the two types of DR profiles for a given mass (\citealt{Dube&Charbonneau2013ApJ}, \citealt{Brun+2021}).\\
Following a similar reasoning, we decided to conserve the sign of $C_s$ when switching the DR regime of the BL model. Indeed, the tilting of emergent structures is thought to be dominated by the Coriolis force over DR sign. So we here suppose that the relative velocity of an emerging flux-tube is dominated by its radial component in the rotating frame. Hence, we considered the influence of $\Omega$ (rotation rate) rather than $\nabla\Omega$ when modeling the effective tilt for the BL effect.
\vspace{1\baselineskip}\\
\begin{figure}
    \centering
    \includegraphics[width=\linewidth]{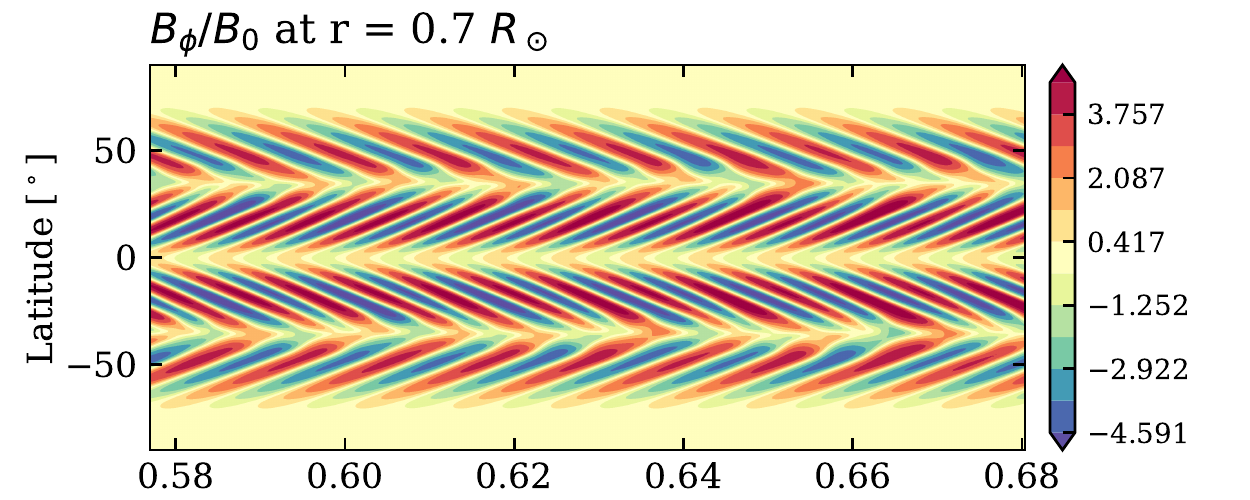}
    \includegraphics[width=\linewidth]{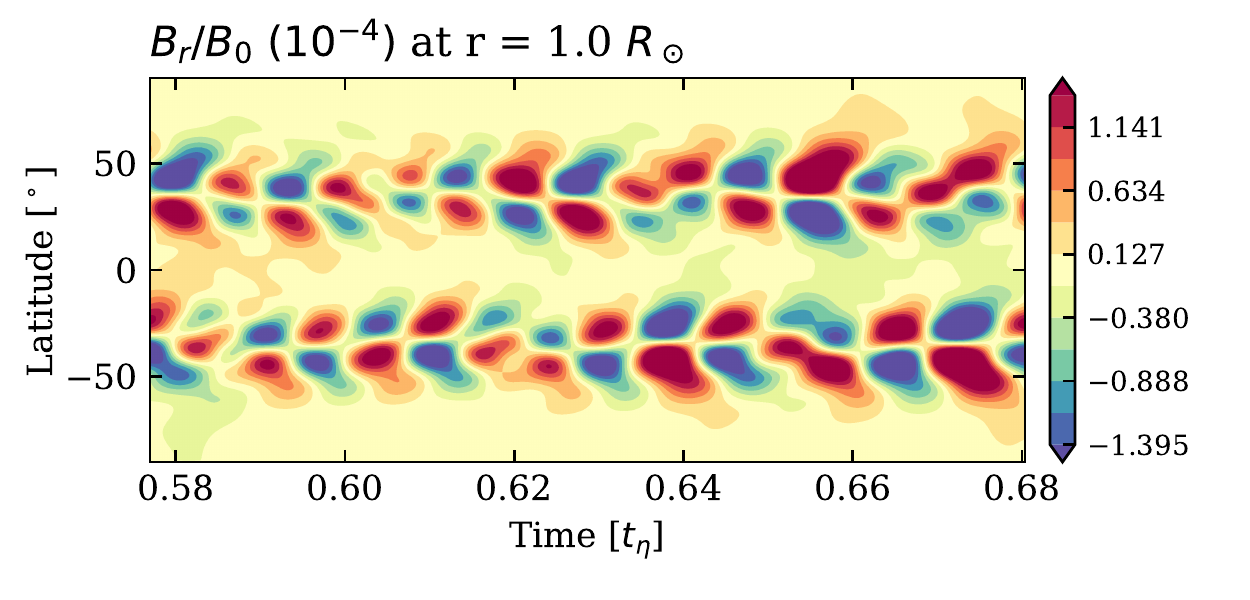}
    \caption{Butterfly diagram of anti-solar $\alpha\Omega$ model with $C_\alpha$=-15 and $D=2.1\times10^6$, for $B_\phi$ at $r=0.7$ and $B_r$ at the surface.}
    \label{fig:Antisolar_AO}
\end{figure}
Butterfly diagrams of the anti-solar $\alpha\Omega$ model are presented in Figure \ref{fig:Antisolar_AO}, where we can observe a magnetic cycle, with new dynamics. First a global convergence toward the co-rotation latitude is observed for the toroidal $B_\phi$ component, shown at the tachocline on the top panel. This was expected, because of the \textit{Parker-Yoshimura} sign rule. It was previously found that wave solutions are allowed in $\alpha\Omega$
mean-field models, first by \citet{Parker1955} for cartesian geometry and then by \citet{Yoshimura1975} for a spherical shape.  Their travel direction, $\bm{s},$ is given by
\begin{equation}
    \label{eq:P-Y prop}
    \bm{s}=\varpi\alpha\nabla\Omega\,\times\,\hat{\bm{e}}_\phi\;\text{and so}\;\left\{
    \begin{array}{ll}
        s_r=\varpi\alpha(\partial_\theta\Omega)/r \\
        s_\theta=-\varpi\alpha\partial_r\Omega\;
    \end{array}
    \right\}
.\end{equation}
In our case, $\partial_\theta\Omega$ and $\partial_r\Omega$ changed sign when we switched to anti-solar DR. That is why we obtained dynamo wave that propagate in the opposite direction to the solar-type case.
Moreover, this observation aptly illustrates the fact that a sign change in both differential rotation and $\alpha$-effect would mathematically makes us reach a mirror solution similar to our solar DR solution (which we checked on models that are not presented here). This can be intuited graphically in Figure \ref{fig:GraphAOBLcomplet} (see the next subsection). The propagation of the polar branches then does not originate from mid-latitudes anymore, where the $\alpha$-effect was more effective for the poloidal generation. They now originate from the poles and the equator, where the $\alpha$-effect is weak. However, we still note toroidal values in both branches similar to the solar reference case. Indeed, the $\alpha$-effect stays strong on the path of equatorial branches, giving more poloidal field, and then toroidal field thanks to the shear. Polar branches stay relatively strong thanks to the radial shear of the tachocline, stronger at high latitudes (see $\partial_r\Omega$ in Figure \ref{fig:AntisolDR}). In addition, we may note that the equatorial branches share now the same origin at the equator, making them phase-locked and of the same magnetic polarity. Hence, a quadrupolar component arises in the magnetic topology at this depth ($r=0.7$ i.e., the tachocline). On the other hand, the polar branch is not phased-locked with the equatorial one of the same hemisphere, as they do not share the same origin of propagation any longer, contrary to the solar DR case. \\
We can next observe the $B_r$ component at the surface, on the bottom panel of Figure \ref{fig:Antisolar_AO}. Despite its fluctuating behavior through different cycles, we see two branches per hemisphere. Their propagation is less clear than deeper down, but they appear to propagate away from each other (poleward and equatorward). The complexity of this dynamics is illustrated in Figure \ref{fig:MeridCuts} with a time series of meridional cuts of magnetic field.
Although this dynamics is regular through different cycles in one hemisphere, both hemispheres are not phase-locked any longer, as previously discussed for the toroidal field at the base of the convection zone. This leads to switches between dipolar and quadrupolar topology represented in Figure \ref{fig:LegAOantisol} with Legendre decomposition odd $l=1,3,5$ and even $2,4,6$ orders, respectively. Moreover, the sign of $B_r$ changes periodically at both poles, leading to global polarity reversals. Its cycle period has nearly halved compared to the solar case, giving now a cycle period of 18 years. We believe this is due to the increasing influence of the symmetric $l=2$ mode that modifies the cycle frequency compared to $l=1$ mode \citep{stixSun2002}.
\begin{figure}
    \centering
    \includegraphics[width=\linewidth]{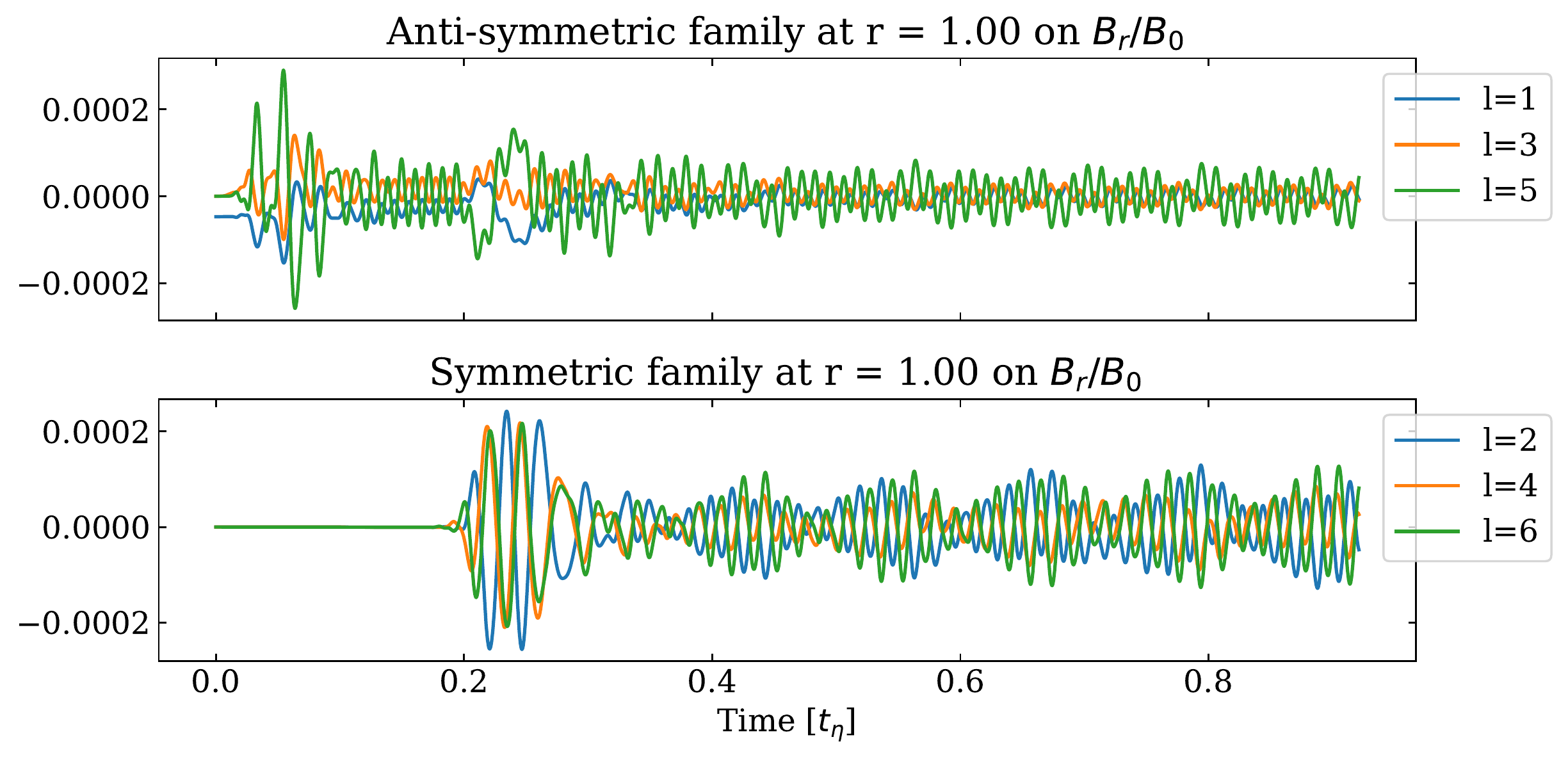}
    \caption{Legendre decomposition on the $B_r$ magnetic component at the surface of the anti-solar $\alpha\Omega$ reference case. Antisymmetric and symmetric families are shown on the top and bottom panel, respectively. The first six polynomial degrees $l$ are shown.}
    \label{fig:LegAOantisol}
\end{figure}
\noindent This halved period seems generic when switching the DR from solar to anti-solar regime, even if we modify the dynamo number $D=C_\alpha.C_\Omega$, as presented in Table \ref{tab:AOstudy}. In both DR regimes, we see the cycle period changing when we modify $D$, but with opposed trends.
The cycle period generally increases with $D$ in the solar like case (except for small $D$), and decreases with $D$ in the anti-solar case. Nevertheless, the cycle period only weakly depends on $C_\alpha$, as shown in Table 2. However, these trends need to be investigated further for low dynamo number $D$.\\
\begin{table}
\centering
\caption{Change of $C_\alpha$, while keeping $C_\Omega=\Omega_{Eq}R_\odot^2/t_\eta=1.4\times10^5$ constant.}
\begin{tabular}{|l|c|c|}
    \hline
    $D\;(10^5)$ & $\odot$ & anti-$\odot$ \\
    \hline
    -7 & 26.2 & 14.1 \\
    \hline
    {\bf -21} & {\bf 31.8} & {\bf 17.7} \\
    \hline
    -42 & 31.4 & 18.8 \\
    \hline
    -63 & 30.7 & 19.5 \\
    \hline
    -84 & 30.0 & 19.9 \\
    \hline
\end{tabular}
\label{tab:AOstudy}
\tablefoot{We report cycle periods in years, considering thus a diffusive time $t_\eta=1550$ years. The period presented are extracted from surface magnetic dynamics. Both reference $\alpha\Omega$ dynamos described in this Section \ref{sec:DR_influence} are shown in bold type.}
\end{table}
We now apply the anti-solar DR profile to the BL flux transport model. The magnetic field is shown in Figure \ref{fig:Antisolar_BL}. First the dynamo becomes stationary without any magnetic cycle, which is in striking contrast to the anti-solar $\alpha\Omega$ case and needs to be investigated more (see next Section). Then we see that the magnetic field is accumulated by the MC at the pole (as represented by black lines) and is dominated by a dipolar configuration. We note that the equilibrium configuration could be either dipolar or quadrupolar depending on the value of $\eta_t$, as was shown in the study of \cite{jouve_role_2007}.
\begin{figure*}
    \centering
    \includegraphics[width=0.75\linewidth]{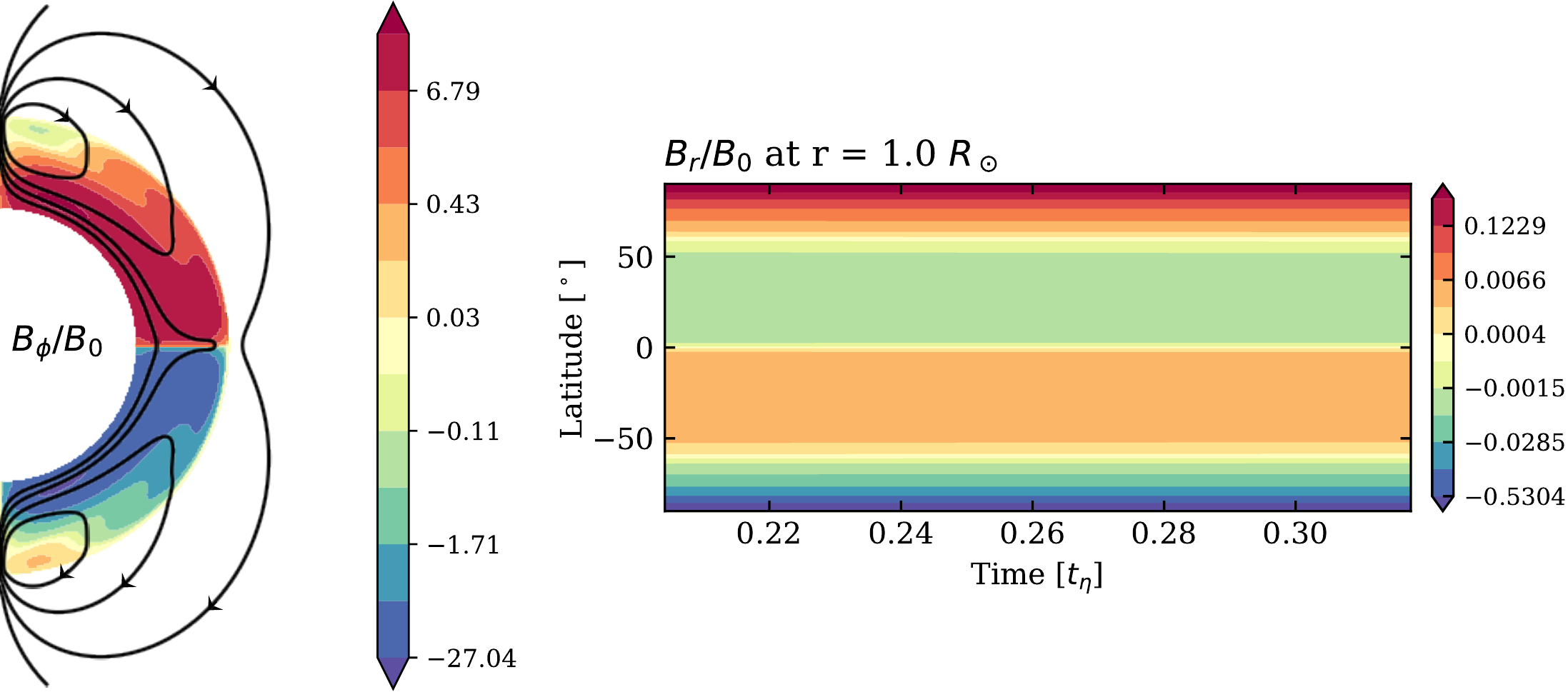}
    \caption{\textbf{Left}: Meridional cut from 0.6 to 1 $R_\odot$, of magnetic components for the reference BL dynamo model in anti-solar DR regime with $C_S=30$ and $D=4.2\times10^6$. Colors are mapped on a logarithmic scale for $B_\phi/B_0$, and black continuous contours are positive $A_\phi$ iso-contours, i.e., $B_{pol}$ clockwise-oriented field lines. Field lines above the surface r = 1$R_\odot$ result from a potential extrapolation. \textbf{Right}: Surface $B_r$ butterfly diagram for the same model, where we clearly see the stationarity of the solution. As in Figure \ref{fig:Sun_BL}, a logarithmic color scale has been chosen.}
    \label{fig:Antisolar_BL}
\end{figure*}

\subsection{Phenomenology of solar and anti-solar dynamos}\label{sec:Phenomeno}
\begin{figure*}
    \centering
    \includegraphics[width=\linewidth]{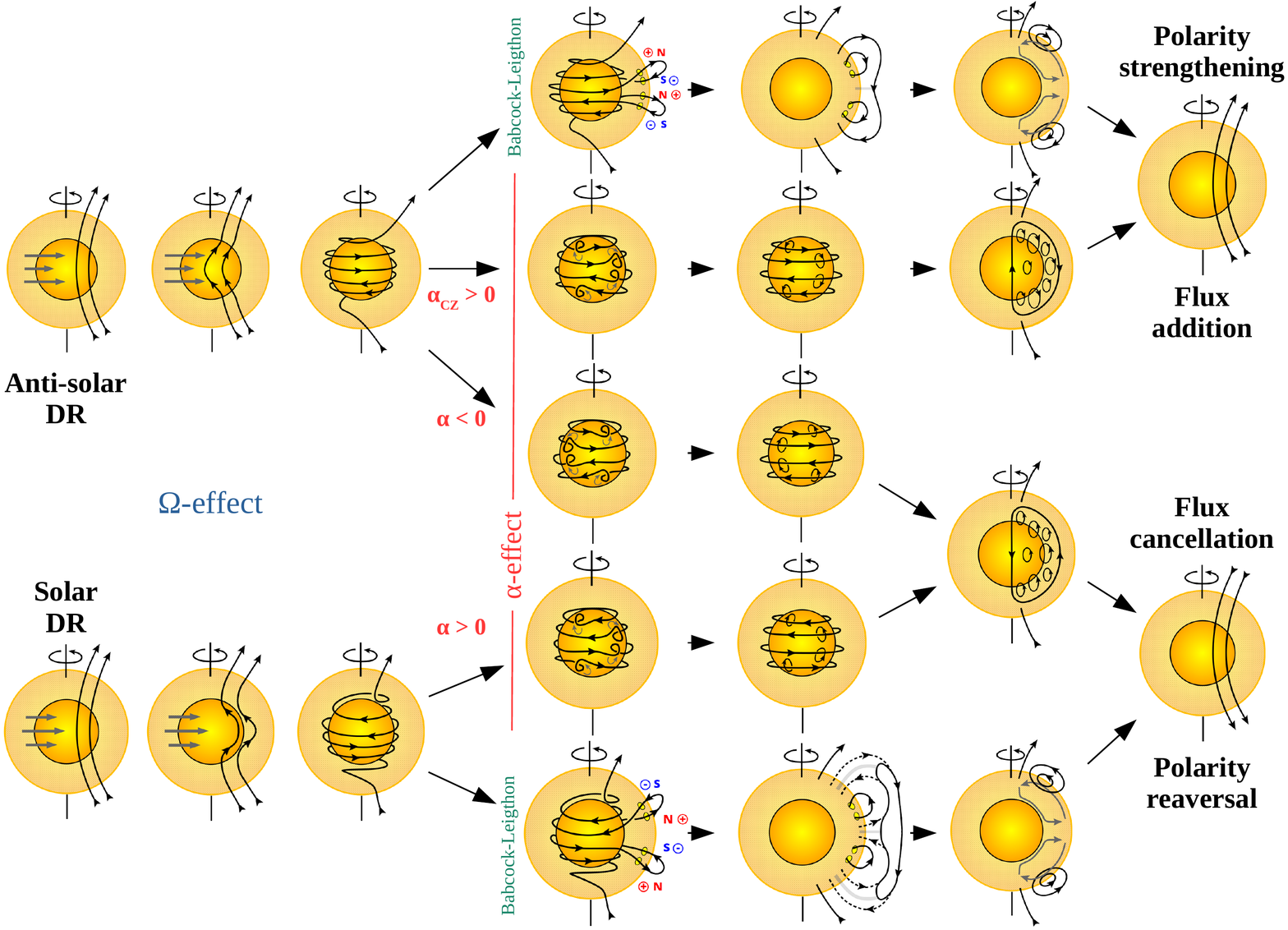}
    \caption{Different steps of geometrical interpretations for $\alpha\Omega$ and BL dynamo models in various configurations. They start on the left side with the differential rotation (DR) shearing magnetic field lines, so-called $\Omega$-effect. It is represented on the first three columns, for anti-solar DR on the top, and the solar one on the bottom (see horizontal grey arrows). The next part (three next columns) illustrates the Babcock-Leighton (BL) mechanism on the first and last rows, while the three middle rows illustrate the $\alpha$-effect. For BL models, we represent the polarity of star-spots (red and blue), current sheet (light grey lines) and the meridional circulation (MC) (darker grey arrows), respectively, in columns 4, 5, and 6. For the $\alpha$-effect, column 4 illustrates orientation of cyclonic motions with grey arrows. Finally, the last columns conclude the dynamo loop, with the presence of a cyclic activity when the final poloidal polarity is opposed to the initial one (bottom), or being stationary otherwise (top). More details are discussed in Section \ref{sec:Phenomeno}. The layout of this Figure is inspired from \cite{sanchezPREDICTABILITYADVECTIONDOMINATEDFLUXTRANSPORT2014}.}
    \label{fig:GraphAOBLcomplet}
\end{figure*}
We can attempt to understand the lost of cycle in the anti-solar DR Babcock-Leighton case by drawing a cartoon (see also \citealt{karak_stellar_2020}). We propose to also include in Figure \ref{fig:GraphAOBLcomplet} case with the $\alpha$-effect, illustrating them for both solar (bottom-left) and anti-solar DR (top-left) regime.\\
We start (in the first column) from global dipole configurations with a similar polarity, inspired by the schematic proposed by \cite{sanchezPREDICTABILITYADVECTIONDOMINATEDFLUXTRANSPORT2014}. On the second column, the DR shears the initial field in different directions according to the DR regime considered: anti-solar (fast poles and slow equator) on the top and solar (slow poles and fast equator) on the bottom. The poloidal field is converted into a toroidal component, whose orientation depends on the DR regime (third column). This mechanism is called the $\Omega$-effect in reference to the shape of sheared poloidal field lines. We then propose on the next three columns different paths in order to restore the poloidal magnetic field from the freshly formed toroidal one.\\
In the first and last rows, we can see how the BL mechanism parametrizes the rise of buoyant magnetic ribbons undergoing the effects of the Coriolis force. These ribbons originate from toroidal magnetic field structures, deeply anchored at the base of the convection zone, and form sunspots when they emerge at the surface, as illustrated on the fourth column. Please note that spots polarities are inverted according to the DR regime considered. The fifth columns shows the diffusion and reconnection of sunspots near the equator. Sunspots closer to the poles diffuse in a poleward branch, reconnecting (or not) according to the polarity of the initial polar field. The freshly formed polar magnetic flux is then transported deeper in the CZ by the MC in column six. This new flux is either added to the initial one in the anti-solar DR regime, or cancels it out in the solar DR regime. In the latter case a global reversal occurs, whereas in the anti-solar case flux can only be added and no reversal occur.\\
The three middle rows illustrate various configurations when considering the $\alpha$-effect. This mechanism parametrizes the cyclonic turbulence of the convection, also influenced by the Coriolis force. Indeed, cyclonic motions can be characterized with the kinetic helicity, experiencing sign changes at the base of the CZ, and can be related to the $\alpha$-effect via Equation \ref{eq:alpha_kin}. Different signs of this effect (here, corresponding to that of $C_\alpha$) correspond to different cyclonic motions (see grey arrows in column four). This effect is named after the shape resulting from the torsion of small magnetic loops. The fifth column illustrates the creation of small-scale poloidal fields, which produces a new large-scale one in average in column six. Finally, this new poloidal flux will either be added to or will cancel the previous one. In the latter case, the cyclic activity can be maintained.\\
Hence, we note that changing the DR is expected to impact strongly dynamo action, likely explaining the results presented in Sections \ref{sec:SuRef} and \ref{sec:AnRef}.
Indeed, these four reference cases correspond to rows 1, 3, 4, and 5 in Figure \ref{fig:GraphAOBLcomplet}. We seek to confirm this interpretation in the next section. To this end, we studied the impact of other characteristics differentiating both types of the dynamo model $\alpha\Omega$/BL, such as the location of the poloidal source term, which will lead us to the second row of Figure \ref{fig:GraphAOBLcomplet}.\\
\section{Robustness of cycles in presence of an anti-solar DR}
\label{sec:RobResults}
We show in the previous section how an $\alpha\Omega$ dynamo in the anti-solar DR regime can allow for a cyclic activity with global polarity reversals, considering deeply located poloidal field sources. We further complete what was considered in previous studies (e.g., \citealt{karak_stellar_2020}). We also note a change of dominant topology between both DR regimes. However, BL flux-transport models seem non-cyclic when an anti-solar DR is considered. In this section, we seek to understand what makes an anti-solar DR dynamo cyclic by exploring fundamental differences between the two anti-solar DR reference models of the previous section.\\
We first study the role of the $\alpha$-effect location. Then we add a meridional circulation (MC) in order to investigate how it modifies the dynamo solutions. Finally, we explore the robustness of the physical interpretation that we proposed in the previous section concerning the stationarity of the anti-solar DR BL dynamo model.

\subsection{Alpha effect location}
\label{sec:alphaLoc}
\begin{figure}
    \centering
    \includegraphics[width=\linewidth]{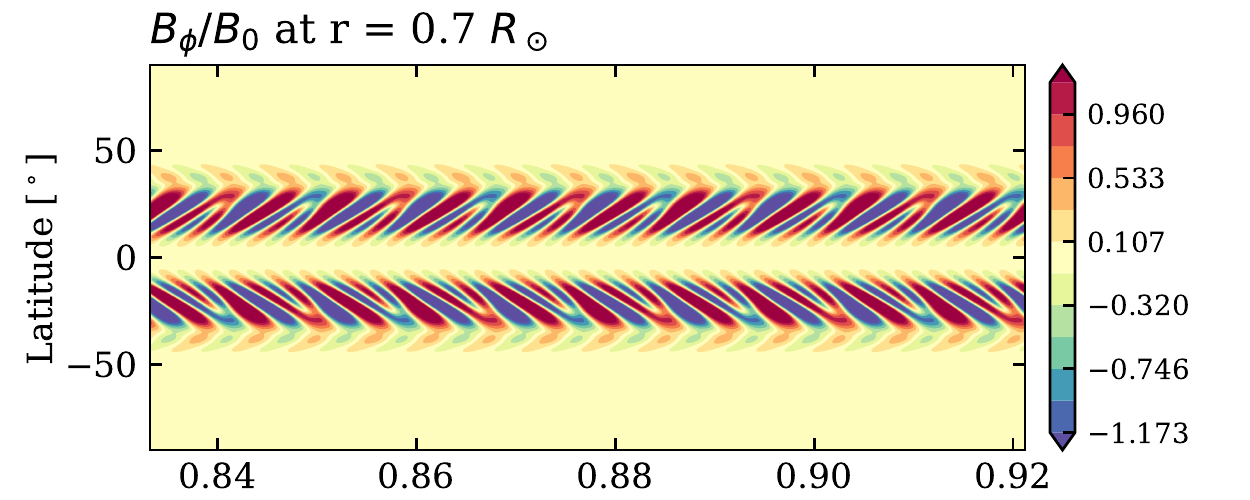}
    \includegraphics[width=\linewidth]{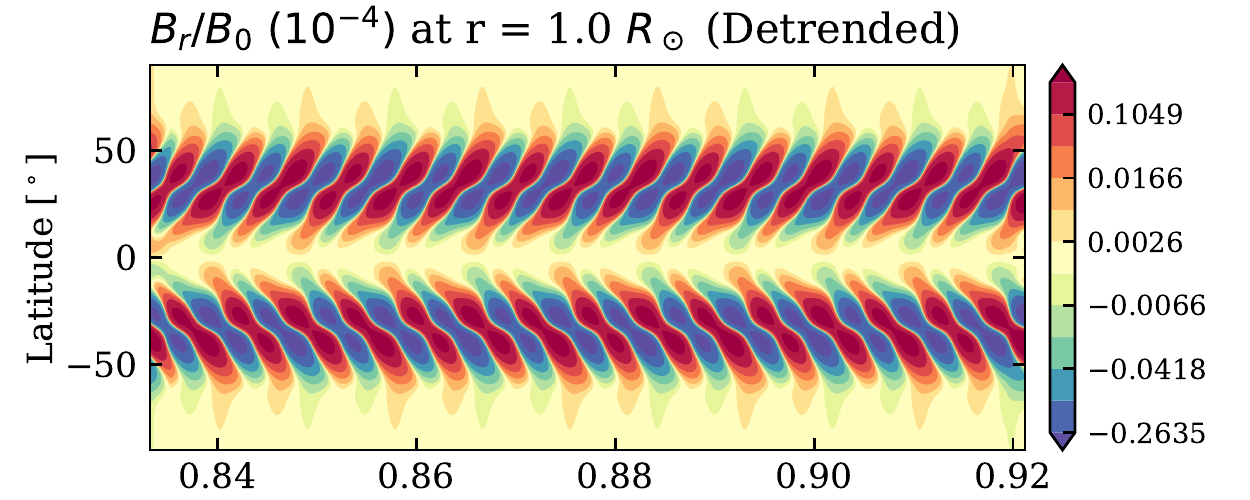}
    \includegraphics[width=\linewidth]{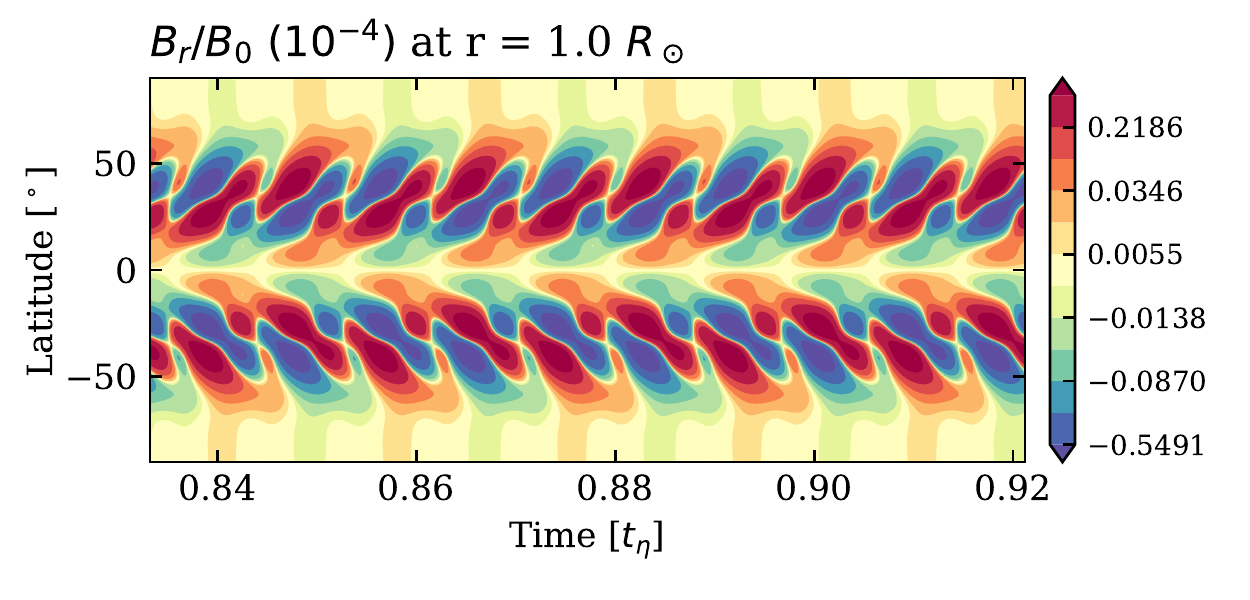}
    \caption{Butterfly diagram of solar $\alpha\Omega$ model at the base of the convection zone with $r_1$=0.7, $r_2$=0.76, $C_\alpha=15,$ and $D=2,1.10^6$. Respectively, $B_\phi$ at $r=0.7$, detrended $B_r$ at the surface and full $B_r$ at the surface (arbitrary units). A logarithmic color scale has been chosen to represent the middle and bottom panels. The middle panel is detrended from the long cycle.}
    \label{fig:Solar776}
\end{figure}

\begin{figure}
    \centering
    \includegraphics[width=\linewidth]{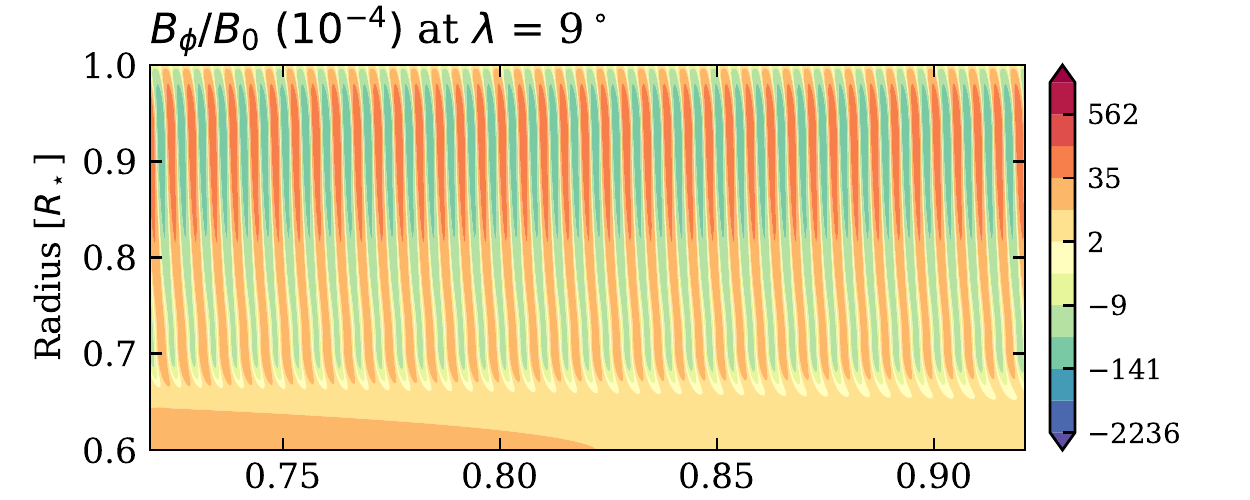}
    \includegraphics[width=\linewidth]{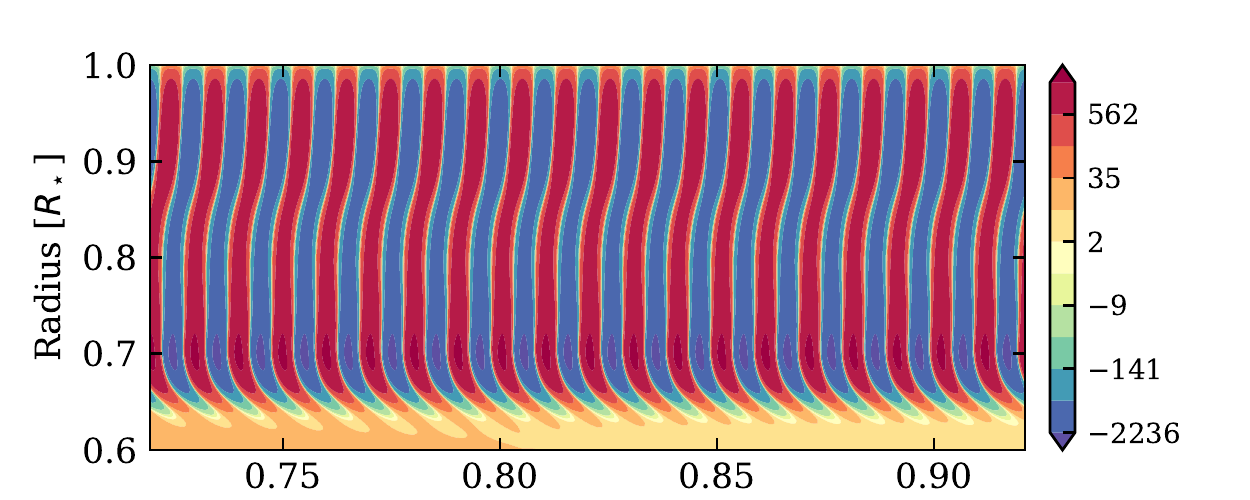}
    \includegraphics[width=\linewidth]{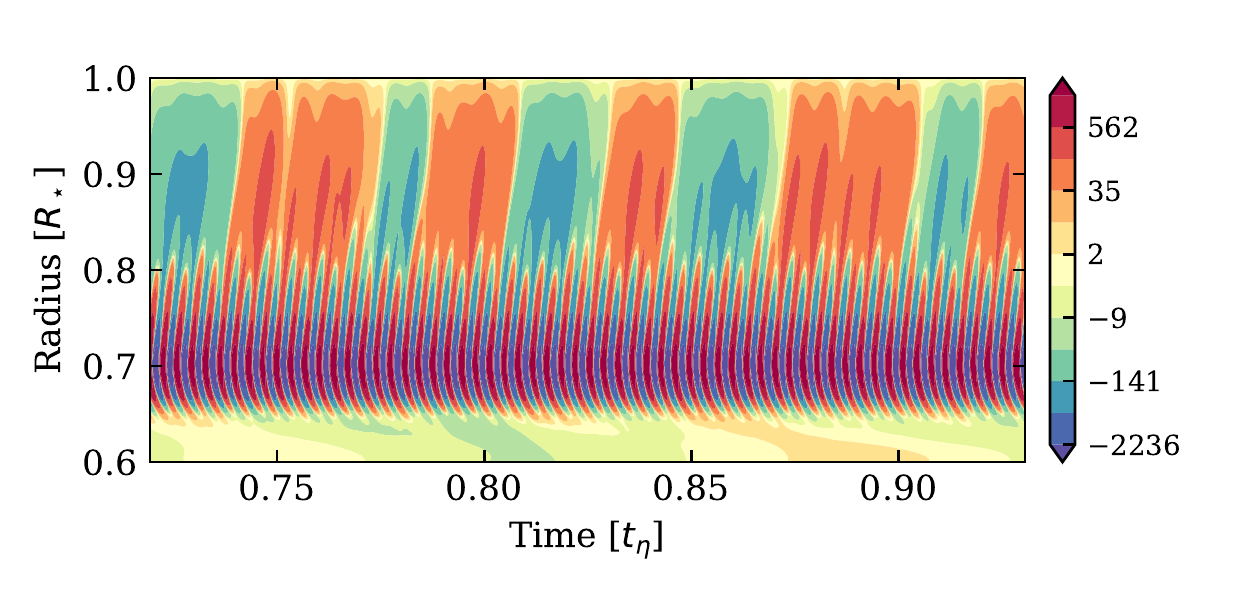}
    \caption{Time-radius diagram of $B_\phi$ in solar $\alpha\Omega$ model with $C_\alpha=15$ and $D=2.1\times10^6$, for models with respectively a location of $\alpha$-term such that $(r_1,r_2)=0.7-0.8$ ; $0.8-0.9$ ; $0.9-1$ (from bottom to top panel). The three are represented on the same radial and temporal ranges, with logarithmic color scales. We note the clear difference of propagation patterns and cycle period, and that the standard deviation of magnetic amplitude increases as the $\alpha$-effect location is deeper.}
    \label{fig:TimeRadBphi}
\end{figure}

One of the main differences between $\alpha\Omega$ and Babcock-Leighton models is the position of the poloidal magnetic field source term (see Figure \ref{fig:phys_ingred}). In this section, we thus explore different locations for the poloidal generation. For this purpose, we considered an $\alpha\Omega$ dynamo model, in which we changed the $\alpha$-effect location using $r_1$ and $r_2$ from Equation \ref{eq:alpha_profile_surf}. An important point is that here we are taking  the opposite sign $C_\alpha=15$ when our $\alpha$-profile is significantly out of the tachocline, within the CZ. As previously discussed the kinetic helicity sign does indeed change at this transition, which directly impacts our $\alpha$-effect according to Equation \ref{eq:alpha_kin}. We offer a systematic presentation of  the models in their solar and anti-solar DR regime for the sake of completeness.\\
We consider various cases of the location of the $\alpha$-effect compared to the reference case: saddled slightly above the tachocline, extended into the CZ, and fully displaced into the CZ. The different runs are summarized in Table \ref{tab:transBush}. In the last row we list the reference $\alpha\Omega$ models with the solar and anti-solar DR, discussed in section \ref{sec:DR_influence}. By keeping an equivalent radial width of the term, we then slightly shift upward its location toward the convection zone. We chose to switch the sign of $C_\alpha$ as soon as $r_1$ is above 0.69$R_\odot$. Cases with no magnetic cycles are labeled with a cross $\times$, otherwise the magnetic cycle period is indicated in years.\\
\begin{table}
\centering
\caption{$\alpha$-effect is distributed through the tachocline and CZ for solar and anti-solar DR regimes.}
\begin{tabular}{|r|l|c|c|}
    \hline
    $C\alpha$ & $r_1$ - $r_2$ & $\odot$ & anti-$\odot$ \\
    \hline
    15 & 0.9 - 1 & 7.3  & $\times$ \\
    \hline
    15 & 0.8 - 1 & 25.1 & $\times$ \\
    \hline
    15 & 0.7 - 1 & 32.4 & $\times$ \\
    \hline
    15 & 0.8 - 0.9 & 15.5 & $\times$ \\
    \hline
    15 & 0.7 - 0.9 & 26.8 (18.9) & $\times$ \\
    \hline
    15 & 0.7 - 0.8 & 64.9 (9.7) & $\times$ \\
    \hline
    \hline
    15 & 0.7 - 0.76 & 27.0 (9.0) & $\times$ \\
    \hline
    15 & 0.69 - 0.75 & 71.5 (10.3) & $\times$ \\
    \hline
    \hline
    -15 & 0.69 - 0.75 & $\times$ & 71.5 (10.3) \\
    \hline
    -15 & 0.68 - 0.74 & 15.8 & 9.0 \\
    \hline
    {\bf -15} & {\bf 0.67 - 0.73} & {\bf 31.8} & {\bf 17.7} \\
    \hline
\end{tabular}
\label{tab:transBush}
\tablefoot{We fix the dynamo number $D=2.1\times10^6$, and magnetic cycle periods are expressed in years, considering a diffusive time, $t_\eta=1550$ years. The period presented are extracted from surface magnetic dynamics. For some cases, a second cycle is present and its period is then indicated in parentheses. Both reference $\alpha\Omega$ dynamos described in the previous section (Section \ref{sec:DR_influence}) are shown in bold type.}
\end{table}
We see that the magnetic cycle disappears in the anti-solar rotation cases for $\alpha$-effects above $r_1=0.69$. It happens when the $\alpha$-source term is no longer spread enough on the tachocline, namely, when the poloidal generation is segregated from the main location of the toroidal one. For the sake of completeness, we added the case where the $\alpha$-effect is confined between $r_1=0.69$ and $r_2=0.75$ for both $C_\alpha$ signs. It illustrates mirrors solutions, where solutions are mathematically switched when we invert one of both dynamo ingredients. The cyclic solutions for anti-solar DR regime is therefore only found when the $\alpha$-effect is saddled on the tachocline, where the radial shear is located. Such solutions therefore exist only in a very narrow parameter space. Meanwhile, the dynamo remains cyclic for solar rotation cases, as we used a classical distributed $\alpha$-effect in our models. It is noteworthy that these models would lose their cycles if $C_\alpha$ is negative and $\alpha$ is segregated enough from the radial shear location, as shown in the last three rows of Table \ref{tab:transBush}.\\
We show in Figure \ref{fig:Solar776} butterfly diagrams when the $\alpha$-effect is at the base of the CZ between $r_1=0.7$ and $r_2=0.76$ for one of the solar models. We clearly see on the top panel the \textit{Parker-Yoshimura} sign rule in toroidal $B_\phi$ dynamics, exhibited poleward along with the equatorward branches. Here the radial expansions of $\alpha$ and $\partial_r\Omega$ indeed still faintly overlap at $r=0.7$. Furthermore, we note that the sign change of $C_\alpha$ induces similar dynamics to the reference anti-solar case shown in Figure \ref{fig:Antisolar_AO}. Comparing the top panels of both models (e.g., Figures \ref{fig:Antisolar_AO} and \ref{fig:Solar776}), we see that the latitudinal extent and the amplitude have decreased; in particular, the strength of the polar branch. Indeed, we see that the $\alpha$-term is concentrated around low latitudes, but now its radial location has been pushed higher in the CZ. The poloidal field production is thus less intense at this radius ($r=0.7$) and has a greater contrast between the polar and equatorial regions. Hence, the toroidal field production is less intense as well, especially at high latitudes. This has a direct consequence on the surface butterfly diagram, where the polarity coming from the pole is now notably less intense, leaving room mainly for the dynamics of the equator polarity going poleward.\\
In addition, we note the presence of a second cycle for some cases where the $\alpha$-effect is located near the bottom of the CZ. We illustrate one of them on the middle panel of Figure \ref{fig:Solar776}, for the case where the $\alpha$-effect is located between $r_1$=0.7 and $r_2$=0.76. In order to highlight such dynamics, the long magnetic cycle has been filtered from the surface butterfly diagram. It appears that such cases are highly sensitive to the numerical resolution, probably due to the particularly sharp profile of the $\alpha$-effect used here (see discussion in Section \ref{sec:ModelDisc}).\\
The upper part of Table \ref{tab:transBush} corresponds to $\alpha$-effects spanning a large part of the convection zone. We first reiterate the absence of magnetic cycle for our anti-solar DR $\alpha\Omega$ cases, even if we increase the value of $C_\alpha$ by an order of magnitude. Dynamos of this type remain stationary and are illustrated on the second row of Figure \ref{fig:GraphAOBLcomplet}. Concerning the solar DR models, we find that the cycle period generally increases as the $\alpha$-effect is located deeper in the CZ and over a larger radial extent.
We illustrate it in Figure \ref{fig:TimeRadBphi} with time-radius traces of $B_\phi$ for 3 models. We note again the \textit{Parker-Yoshimura} sign rule, this time with dynamo waves propagating radially outward, thanks to the latitudinal shear $\partial_\theta\Omega$ (see Equation \ref{eq:P-Y prop}). Indeed, the $\alpha$-effect is now in the bulk of the CZ, where $B_\phi$ is generated only by the latitudinal shear, as there is no radial shear there.\\
Finally, we note again the presence of a shorter additional cycle in the model with the $\alpha$-effect located between $r_1$=0.7, $r_2$=0.8.\vspace{1\baselineskip}\\
To summarize, we see that the magnetic cycle period in $\alpha\Omega$ dynamos depends on the location and thickness of our $\alpha$-source term. Moreover, we note in anti-solar DR regime that moving this poloidal source term away from the tachocline is enough to suppress the magnetic cycle. So the disappearance of cyclic behavior in our reference BL model (Section \ref{sec:AnRef}) is likely linked to the radial separation between both toroidal and poloidal sources. However, this is not the only particularity of such flux-transport models. The meridional circulation plays a key role, linking dynamically the two sources of the magnetic fields. In the next section, we propose to study the effect of such a velocity field on $\alpha\Omega$ models, in particular the anti-solar ones.\\

\subsection{Impact of the meridional circulation on $\alpha\Omega$ dynamo models}\label{sec:MC}
\begin{figure}
    \centering
    \includegraphics[width=\linewidth]{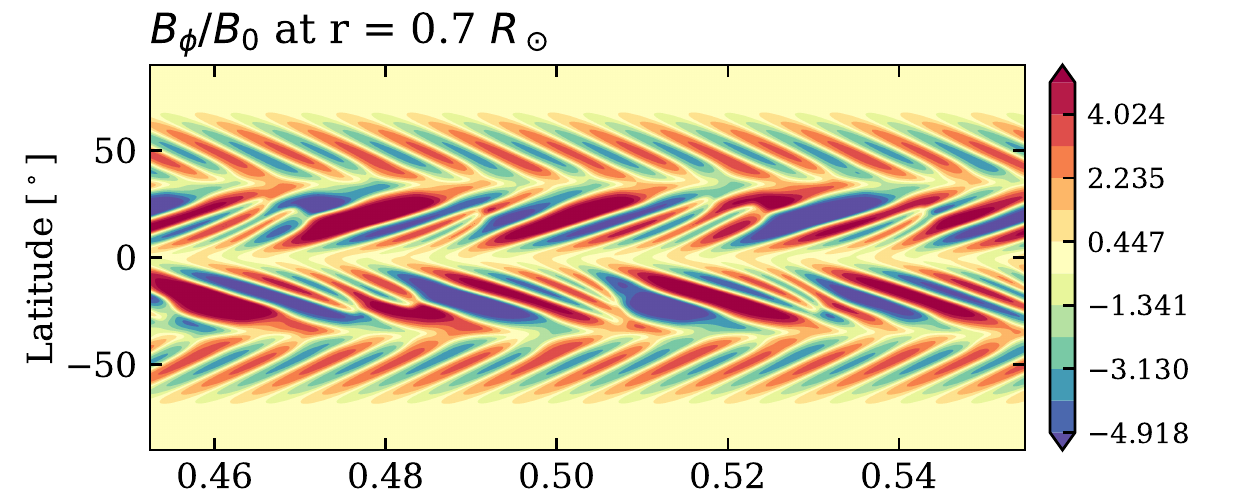}
    \includegraphics[width=\linewidth]{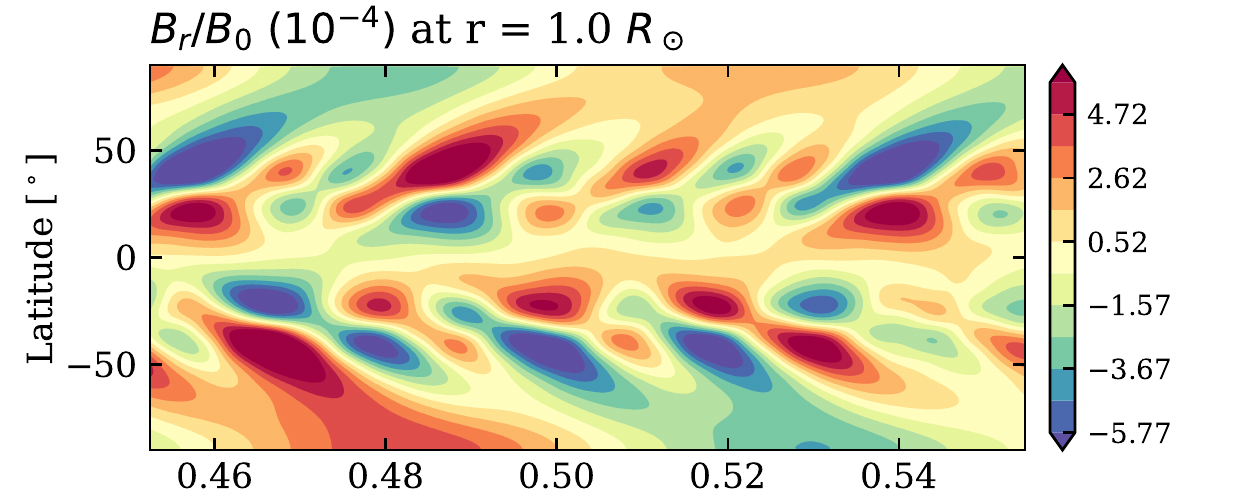}
    \includegraphics[width=\linewidth]{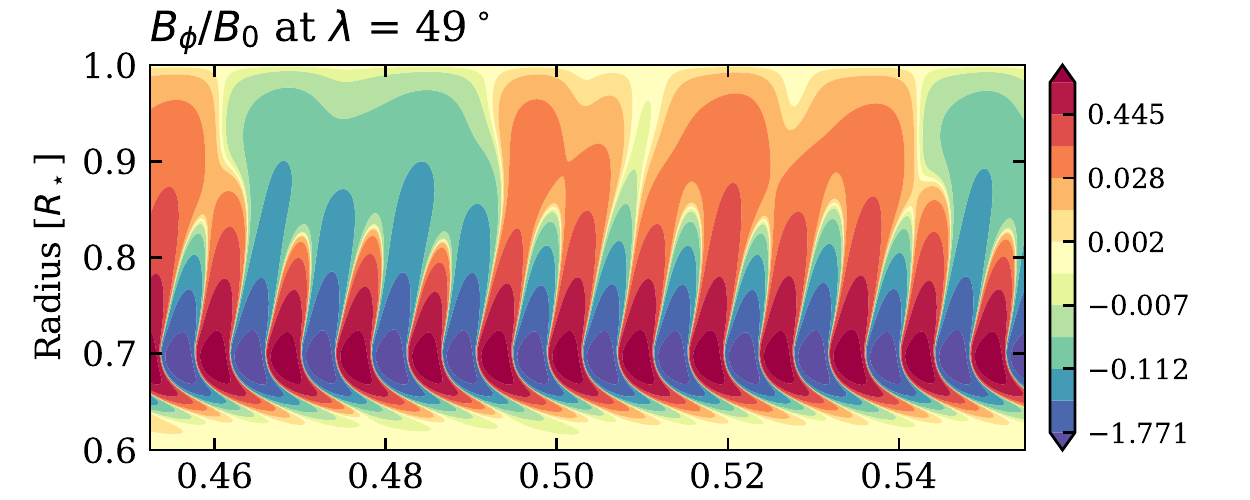}
    \includegraphics[width=\linewidth]{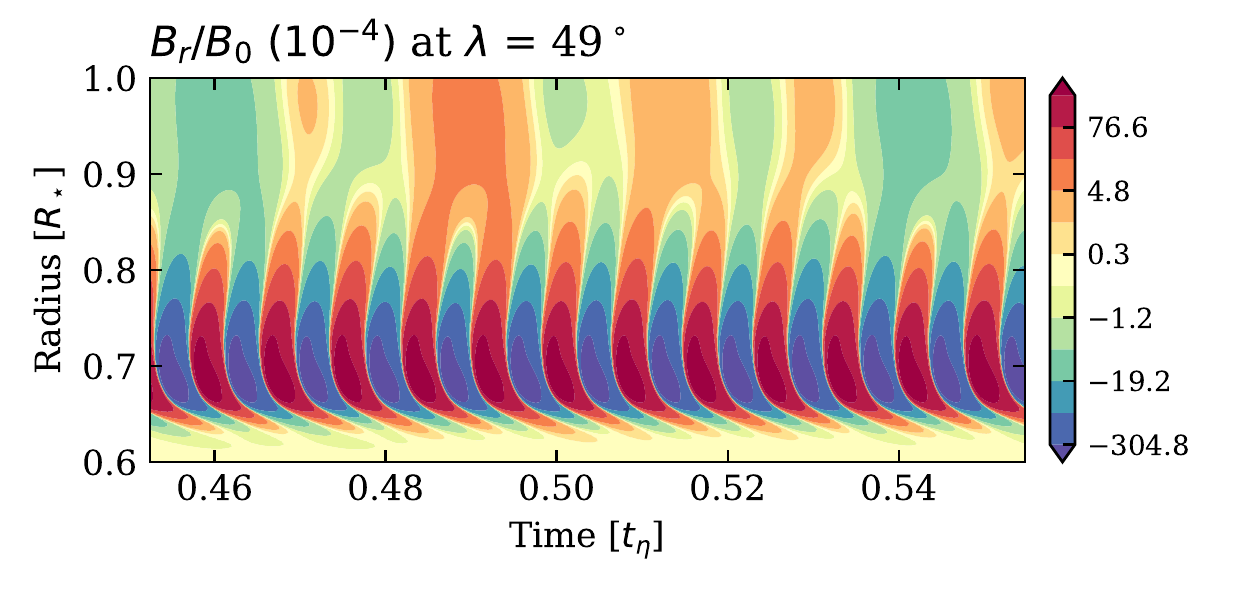}
    \caption{Both top panels are butterfly diagrams of the anti-solar $\alpha\Omega$ model with $C_\alpha$=-15, $D=2.1\times10^6$, where we added a MC with $R_{\rm m}=50$. The $\alpha$-source term is at the base of the convection zone with $r_1$=0.67 and $r_2$=0.73. Respectively, $B_\phi$ at $r=0.7$ and $B_r$ at the surface. Both bottom panels represent time-radius plots, respectively for $B_\phi$ and $B_r$ at $\lambda=45^\circ$. Logarithmic color scales are chosen for second and last panels.}
    \label{fig:refAnMC}
\end{figure}

Observations of the Sun have revealed large-scale flows on the solar surface. In particular meridional flows from the equator toward the pole have been found at the surface, which may be  evidence of a global meridional circulation in the convection zone, as illustrated in panel (c) of Figure \ref{fig:phys_ingred}. Although its existence is well established down to $0.9\;R_\odot$, disagreement still exist about what occurs deeper down (\citealt{gilesSubsurfaceFlowMaterial1997}, \citealt{zhaoDETECTIONEQUATORWARDMERIDIONAL2013}, \citealt{Basu2020}).\\
As described in Section \ref{sec:PolGene}, MC is considered essential ingredient in the study of stellar magnetism dynamics in flux-transport models (\citealt{choudhuriSchusslerDikpati1995}, \citealt{jouve_role_2007} and \citealt{karakIMPORTANCEMERIDIONALCIRCULATION2010}). It is also a large-scale flow genuinely emerging in global 3D non-linear numerical simulations (\citet{Featherstone2015}, \citet{Brun+2017}). Therefore, taking into account this physical ingredient is crucial in understanding the role of large-scale flows in the resulting dynamics.\vspace{1\baselineskip}\\
The effects of the MC on $\alpha^2\Omega$ dynamos have been studied by \citet{Kuker+2001} and \citeauthor{Bonanno2002} (\citeyear{Bonanno2002}, \citeyear{Bonanno+2003}). Following their lead, we here propose to consider similar models than in the previous section, shifting the poloidal field generation from the tachocline to the surface. As discussed earlier in this paper, we switched the sign of $C_\alpha$ according to the $\alpha$-term location. In addition, we now have to consider the magnetic Reynolds number characterizing the MC. We considered $R_{\rm m}=500,$ as in the BL models described in Section \ref{sec:DR_influence}.\\
Here, we present some of the results of these experiments (Table \ref{tab:BushSurfMC}). First, we note that for both DR regimes, the meridional circulation makes it harder to maintain a cycle, as there are fewer cyclic models available for this parameter range. Once more, they are totally absent for the anti-solar regime when the poloidal generation is located in the bulk of the CZ, as expected based on Figure \ref{fig:GraphAOBLcomplet}. Additionally, we observe that the presence of the MC makes cyclic activity disappear when considering $\alpha$-effect locations near the surface. This might result from destructive interference, occurring in the polar dynamo branch, due to strong advection when the $R_{\rm m}$ becomes high enough. Obviously, decreasing the MC amplitude is then one solution for recovering the polarity cycle. On the other hand, an increase by an order of magnitude of the $\alpha$-effect amplitude helps to recover the cycle. It shows the subtle balance between both $\alpha$-effect and MC, which determines which mechanism dominates the other. The model can be indeed significantly impacted by the MC when its amplitude, $u_0=R_{\rm m}\eta_t/R_\odot$, becomes comparable to the characteristic propagation speed of the dynamo waves.\\
When solar models remain cyclic, some cycle periods are shortened by the MC. Considering reasonable $R_{\rm m}$ values, this can be understood as the meridional flow accelerating the migration of newly-generated toroidal magnetic field toward the surface source term. However, we show in Table \ref{tab:BushSurfMC} that there is no general trend and that the period may also increase. It indeed illustrates this subtle balance of the MC effect and the $\alpha$-generation well, as shown by \citet{RobertsStix1972} or \cite{Yeates2008}.\\
\begin{table}
\centering
\caption{$\alpha$-effect from the tachocline, through the convection zone, with a meridional circulation, $Rm=500$ (50 for starred $^*$ case), for solar and anti-solar DR.}
\begin{tabular}{|r|l|c|c|}
    \hline
    $C_\alpha$ & $r_1$ - $r_2$ & $\odot$ & anti-$\odot$ \\
    \hline
    15 & 0.95 - 1 & $\times$  & $\times$ \\
    \hline
    15 & 0.9 - 1 & $\times$  & $\times$ \\
    \hline
    15 & 0.8 - 1 & 21.1 & $\times$ \\
    \hline
    15 & 0.7 - 1 & 24.0 & $\times$ \\
    \hline
    15 & 0.8 - 0.9 & 23.4 & $\times$ \\
    \hline
    \hline
    -15 & 0.67 - 0.73 & $\times$ & $\times$ \\
    \hline
    -15$^*$ & 0.67 - 0.73 & 24.4 & 121.8 (31.3)\\
    \hline
\end{tabular}
\label{tab:BushSurfMC}
\tablefoot{Magnetic cycle periods are expressed in years, considering a diffusive time $t_\eta=1550$ years. The period presented are extracted from surface magnetic dynamics.}
\end{table}
We then applied a MC to our $\alpha\Omega$ reference cases (from Section \ref{sec:AnRef}). The addition of this large-scale motion makes their cyclic activity disappear. As discussed earlier in this work, this is likely due to destructive interference from the strong advection of the MC. Cycles can be recovered by decreasing the MC amplitude, as shown on the last row of Table \ref{tab:BushSurfMC}. This happens in this model for $R_{\rm m}\sim50$ and we propose to illustrate it in Figure \ref{fig:refAnMC}. We now clearly see the signature of the MC with an equatorward toroidal flux accumulation at the base of the CZ (top panel). The equatorial branch going poleward is now propagating slower, compared to the reference case in Figure \ref{fig:Antisolar_AO}, while the polar branch propagates even quicker. Indeed, the action of the MC shortens locally the cyclic period of polar structures to 13 years, while it locally extends  the cycle period of the equatorial dynamics to 31 years. This equatorward advection then accumulates more magnetic field in the equatorial branch. Equatorial cyclic patterns then rise toward the surface, where they govern the dynamics, as shown on the second panel. Time-radius panels confirm that indeed the short cycle remains confined to the base of the CZ for these low latitudes. Finally, we note that this 31.3-year cycle is not perfectly regular, which results in an even longer cycle at the surface in polar regions. Indeed, the global dipole polarity is reversed with a period of 121.8 years. When $R_{\rm m}$ increases, this long cycle period increases, and the cycle ultimately disappears at large $R_{\rm m}$. This is illustrated in the second to last row of Table \ref{tab:BushSurfMC}, where the dynamo is stationary. We shown in the next subsection how adding a surface source term impacts this model.\vspace{1\baselineskip}\\
In the end, adding a MC to $\alpha\Omega$ models generally does not help to obtain a magnetic cycle if one did not exist before, regardless of the DR regime. Finally, studying the location of the poloidal source term in a flux transport model leads us to conclude that it is even more difficult to obtain a magnetic cycle the closer this term is to the surface. In particular, this allows us to conclude that the magnetic cycle loss of the anti-solar BL model is not caused by its non-local character, but rather by an unfavorable field polarity interplay.

\subsection{Considering whether a Babcock-Leighton dynamo with an anti-solar DR could sustain a magnetic cycle}
\label{sec:hybrid}
\begin{figure}
    \centering
    \includegraphics[width=\linewidth]{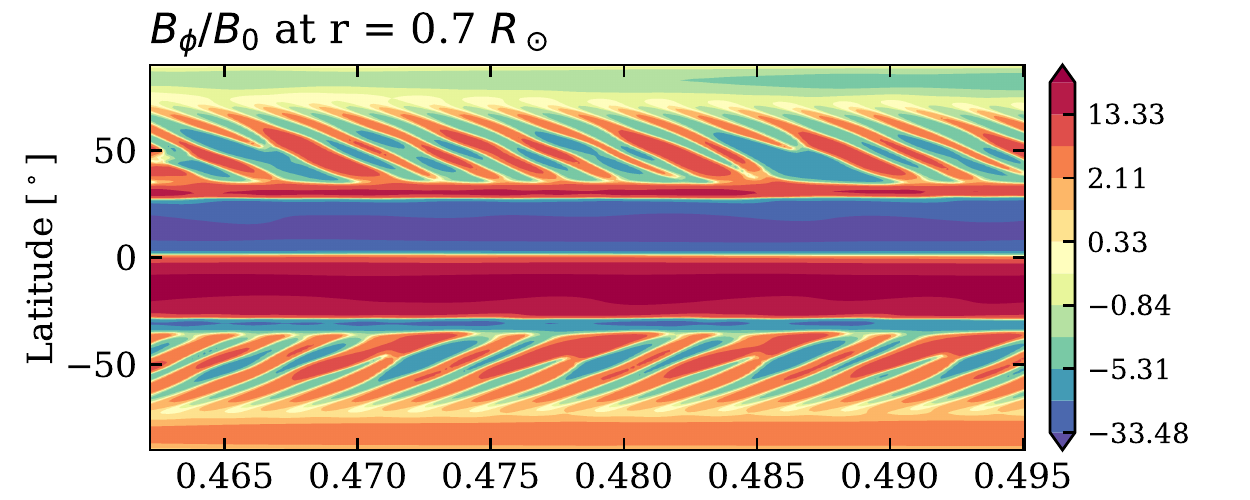}
    \includegraphics[width=\linewidth]{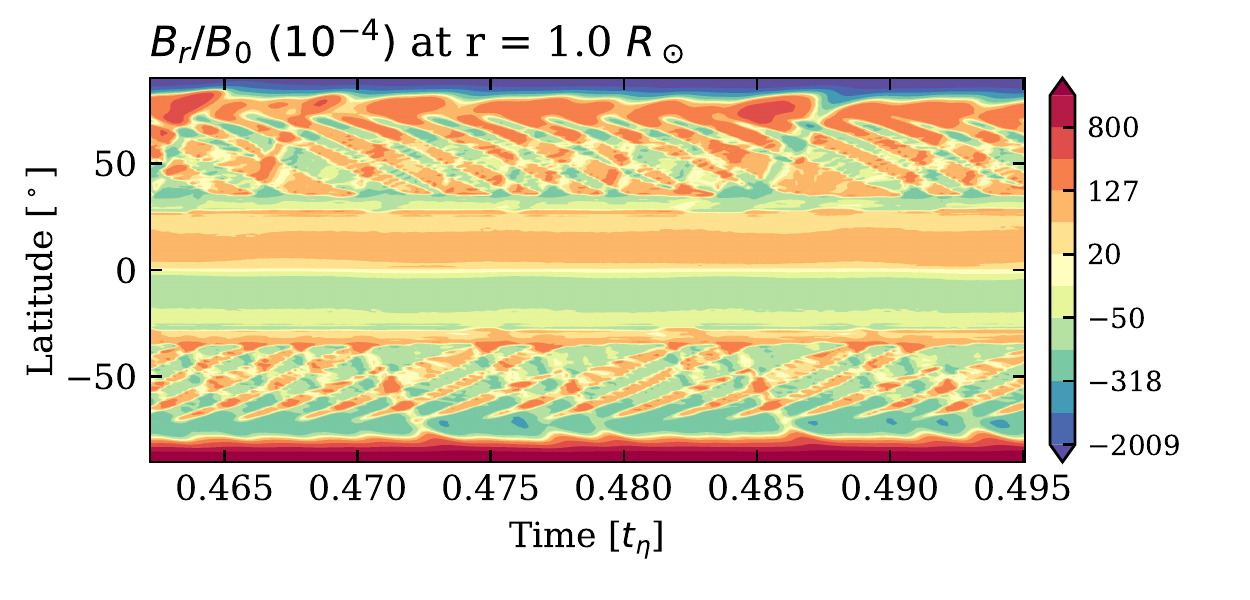}
    \caption{Butterfly diagram of solar BL flux-transport model with $C_S$=30 and $R_{\rm m}=500$, where an $\alpha$-source term is added. This term is saddled on the tachocline between $r_1$=0.67 and $r_2$=0.73, with $C_\alpha=-150$. Respectively, $B_\phi$ at $r=0.7$ and $B_r$ at the surface. A logarithmic color scale has been chosen to represent the bottom panel.}
    \label{fig:AO150}
\end{figure}

\begin{figure}
    \centering
    \includegraphics[width=\linewidth]{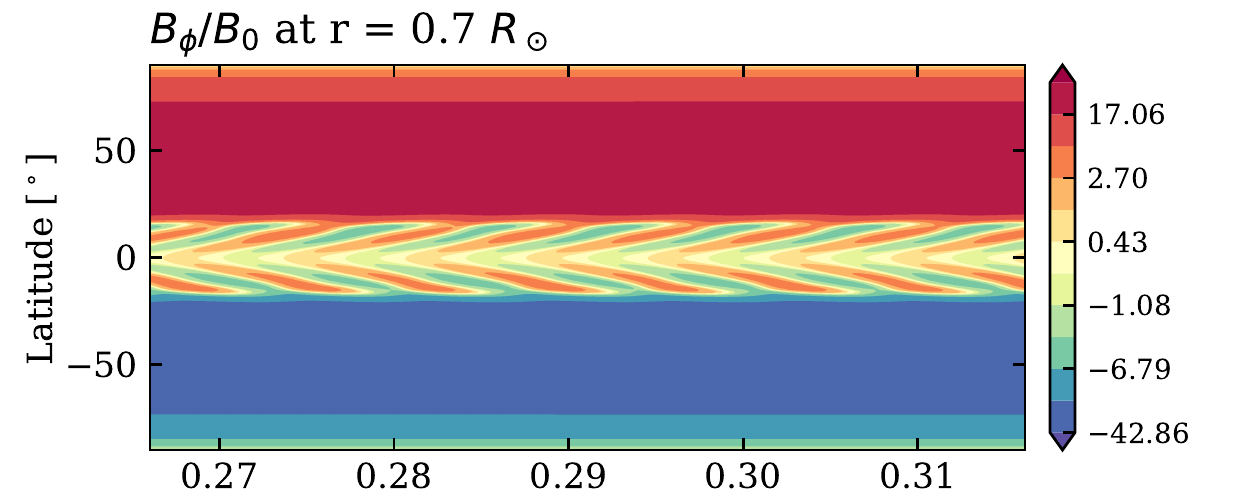}
    \includegraphics[width=\linewidth]{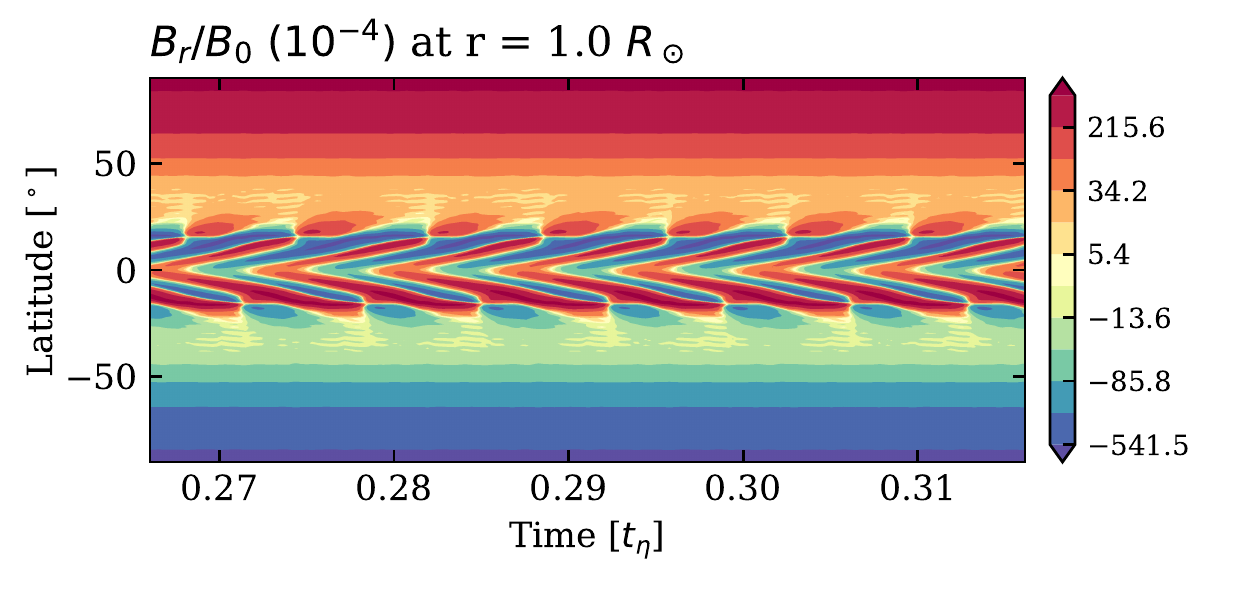}
    \caption{Butterfly diagram of solar BL flux-transport model with $C_S$=30 and $R_{\rm m}=10$, where an $\alpha$-source term is added. This term is saddled on the tachocline between $r_1$=0.67 and $r_2$=0.73, with $C_\alpha=-15$. Respectively, $B_\phi$ at $r=0.7$ and $B_r$ at the surface. A logarithmic color scale has been chosen to represent both panels.}
    \label{fig:hybridRm10}
\end{figure}

As we show in Section \ref{sec:DR_influence}, the magnetic cycle disappears from our BL model when we impose an anti-solar DR regime and, therefore, the dynamo becomes stationary. We can see in previous subsections that this disappearance is likely caused by the location of the BL source term. The interplay between polarities could then lead to a flux addition, making  global topology reversals impossible, as illustrated in Figure \ref{fig:GraphAOBLcomplet}. The goal of this section is, thus, to test the robustness of such a conclusion.\\
To this end, we have carried out some exploration of the reference BL model's parameters space, investigating the different factors. The $C_\Omega$ value has been decreased by up to a factor of 5, motivated by a stellar rotation rate, which is likely to be significantly lower in anti-solar DR regimes. The amplitude of the shear $\Delta\Omega/\Omega_{\rm Eq}\sim26\%$ has also been decreased up to $10$ and $1\%$ by changing coefficient of Table \ref{tab:DR_coefs}. Only the switch toward solar DR can bring the cyclic activity back. The impact of a double bump in the $\eta$ profile (see comment and references at the end of Section \ref{sec:SuRef}) has been explored as well. We changed the amplitude of the MC, decreasing it until 0, considering even clockwise circulation in the northern hemisphere (negative values for $R_{\rm m}$). But as expected earlier, no magnetic cycle has been seen in our flux-transport BL model as soon as the anti-solar DR profile was set. All corresponding butterfly diagrams stay equivalent to the one presented in Figure \ref{fig:Antisolar_BL}.\vspace{1\baselineskip}\\
We try subsequently to add in this model the $\alpha$-profile providing a cycle in the anti-solar $\alpha\Omega$ model, from Section \ref{sec:AnRef}. We therefore couple the anti-solar BL-flux transport model to an additional $\alpha$-effect located between $r_1-r_2=0.67-0.73$. Although the dynamo remains stationary with $C_\alpha=-15$, we need to decrease $C_\alpha$ below $-70$ in order to obtain a cyclic activity. We illustrate such a model in Figure \ref{fig:AO150}, where $C_\alpha=-150$. The top panel shows us the signature of the MC, with a strong equatorward accumulation of toroidal field in the equatorial region at the bottom of the CZ. We also note clear polarity inversions located at mid-latitudes at the tachocline and the surface, without polarity reversals of the global poloidal field. We also find very few changes in the cycle period with regard to $C_\alpha$ (based on models not included in this paper). On the other hand, the cycle period has been found to be mainly controlled by the MC dynamics in a advection-dominated regime.\\
In that sense, another way to obtain a cyclic behavior, while keeping $C_\alpha=-15$, is to decrease the MC amplitude. The dynamo then becomes cyclic when we go below $R_{\rm m}\sim25$. We illustrate such a model for $R_{\rm m}=10$ in Figure \ref{fig:hybridRm10}, where we note again local polarity reversals. As a result of a weaker MC, we see on the top panel that they are now localized around the equator and emerging at the surface as shown in the bottom panel. Coming back to the schematic in Figure \ref{fig:GraphAOBLcomplet}, the polar dynamics can be interpreted here as the BL dynamo path on the first row, while the cyclic local dynamo around the equator can be illustrated with the third row.\\
We can conclude that in presence of the MC with an anti-solar DR, a BL source-term only will not be able to maintain a cyclic activity. However, an $\alpha$-effect can create global polarity reversals in the presence of a MC, but only if it is the only source of poloidal magnetic field, which then has to be located at the bottom of the convection zone. Indeed, we observe that global polarity reversals disappear at both poles when combining both $\alpha$-effect and BL-effect. Only a local cyclic dynamo near the equator is preserved in this hybrid case. Ultimately, we find that the MC can prevent a cyclic behavior when it is fast enough (i.e., when the Rm is large enough).

\section{Context and discussions}\label{sec:Context}

\subsection{Astrophysical context}\label{sec:AstroCont}
The STELEM code used in this study computes mean-field equations in a dimensionless form. From a dynamical system point of view our interest was focused on keeping similar dynamo numbers, $D$, in order to conserve all our reference models in the same dynamical range. This explains why we do not have the exact solar magnetic cycle period for the reference $\alpha\Omega$ model of Section \ref{sec:SuRef}.\\
Now taking the astrophysical point of view, we could relax the constraint on keeping $D$ similar between models and fine-tune the solar period. This can be done by increasing the turbulent diffusivity $\eta_t$ through a decrease of $C_\Omega$ value. To this end, we can get the 22-year magnetic cycle period of our Sun, from the $\alpha\Omega$ model of Section \ref{sec:SuRef}, by decreasing $C_\Omega$ to $9.3\times10^4$. We thus still consider $\Omega_{Eq}/2\pi=460$ nHz at its solar value, and keep $C_\alpha=-15$, which leads to new values for $t_\eta=1030$ years, $\eta_t=1.5\times10^{11}$ cm$^2$.s$^{-1}$ and $\alpha_0=33.2$ cm.s$^{-1}$. We illustrate this model in Figure \ref{fig:SunModelAOtuned} of the Appendix. Despite a better agreement with the observed magnetic solar period, the dynamical solution is highly similar to the reference model of Section \ref{sec:SuRef}, which confirms the interest of this study from the point of view of dynamical systems.\\
Moreover, global 3D numerical simulations of stellar convection show that anti-solar differential rotation (fast poles - slow equator) might emerge for high Rossby numbers with a transition around the unity. Hence, this DR regime is likely to occur for slow rotators, when considering equivalent convective characteristics. Following the same astrophysical point of view, we can adapt the rotation rate with lower $C_\Omega$. We note that thanks to the dimensionless form of the equations, the reference anti-solar DR $\alpha\Omega$ model in Section \ref{sec:AnRef} can also be seen as a star rotating slower than the Sun and less diffusive than $\eta_t=10^{11}$ cm$^2$.s$^{-1}$, while conserving $C_\Omega=1.4\times10^5$. Again, this approach leads to the same conclusions as the ones presented in this work.\\

\subsection{Model context}\label{sec:ModelDisc}
For the sake of completeness, several points have to be addressed in order to present the context of our model choices and how to improve them in future work.\\
First, the analytical fit used in Equation \ref{eq:Thomson_DR} captures most of the solar internal rotation properties, but does not account for radial variations within the convection zone. In particular, we do not take into account the near-surface  shear layer (NSSL). Such a layer has been shown to have a potential impact on dynamo migration \citep{karakBABCOCKLEIGHTONSOLAR2016}, and is particularly interesting in terms of the Babcock-Leighton dynamos as it introduces a shear near the location of poloidal field generation. Nevertheless, we do not yet know  whether slow rotator with anti-solar DR would go on to develop an NSSL-like layer, thus, we opted to omit this aspect in the present study. We shall consider its inclusion in a future work.\\
Second, Equation \ref{eq:alpha_kin} gives us an expression of the parametrization of turbulence by the parameter $\alpha$ in the mean-field theory context. We chose to conserve $C_\alpha$ sign while transiting the DR regime as the kinetic helicity is likely to do as well. Nevertheless, this expression has several forms, especially if  a current helicity term is meant to be taken into account, such that $\alpha = -\frac{\tau_C}{3} (\bm{u'} \cdot (\nabla \times \bm{u'}) - \frac{1}{c\rho} \bm{j'} \cdot \bm{b'})$; as introduced by \cite{Pouquet+1976}, discussed in \cite{BrandenburgSubramanian2005}, and illustrated in stellar convective shells of \cite{Warnecke+2018}. Recent global MHD numerical simulations of \cite{Brun+2021} do not show any significant sign or behavior change of the current helicity profile between the different DR regimes.\\
Then multiple choices of meridional flow can be made in order to link toroidal and poloidal flux. As the amplitude of the solar MC is relatively low (observed around $10-20$m.s$^{-1}$ at the surface, and thought to be $\sim4$m.s$^{-1}$ at the base of the CZ), its inversion from helioseismology can be very difficult. Indeed, some observers detect multiple cells per hemisphere, while others do not (see \citealt{Gizon+2020} and references in Section 5.4.3 of \citealt{Charbonneau2020LRSP}). In the meantime, global numerical simulations of stellar convection tend to predict multiple cells for solar DR against only one per hemisphere for the anti-solar regime (see \citealt{Brun+2017}). Thus, this physical process is currently subject to debate and we refer interested readers to \citet{Featherstone2015} for a detailed theoretical study. Here, we chose a single cell hemispheric meridional flow to follow our objective of maintaining simplicity. Finally, other choices could have been made in order to link magnetic fluxes, such as diffusion-dominated model or turbulent pumping (see, respectively, \citealt{Yeates2008} and \citealt{DoCao2011}). Such considerations come out of the scope of the present study.\\
We observe in Section \ref{sec:alphaLoc} that second cycles with shorter periods can emerge in some cases when increasing the resolution of the numerical simulation. It happens near the tachocline, when the transition between stationary and cyclic dynamos appears for the anti-solar DR regime.
It seems that the period of such models is very sensitive to the resolution, which is likely due to the sharpness of the chosen $\alpha$-profile. Nevertheless, numerical resolutions do not change the nature of our solution (cyclic or stationary) and, therefore, they do not affect the main conclusions of this study. This is the main reason why we report dynamo models with a resolution of 128 x 256, namely, to ensure a good solution convergence (as in the benchmark published in \citealt{jouve_solar_2008}).\\
Next, deciding on which formulation to take for the $\alpha$-effect is not an easy task, as its profile is not simple to determine exactly, even from 3D simulations where all the 3D information is known (see \citealt{SimardCharbonneauDube2016} for instance). We ultimately chose to use a similar profile to the one used in \cite{Bushby2006}, as it allows us to reproduce more faithfully the surface magnetic field of the Sun through simulation and to then explore which region is important for cyclic activity.\\
Moreover, we made the choice of an isotropic $\alpha$-effect for the sake of simplicity regarding the interpretation of our results. Indeed, it allows us to highlight quite clearly the role of each chosen ingredient and how they act in an anti-solar DR regime. Some studies seem to extract significantly anisotropic $\alpha$ tensors for fast rotators; however, this is not the case for slow rotators, where we are likely to find an anti-solar DR profile (\citealt{WarneckeKapyla2020}).\\
With the same goal of maintaining simplicity, we decided to use $\alpha\Omega$ models instead of $\alpha^2\Omega$. Indeed, we assumed in these cases that a toroidal $\alpha$-effect would be negligible in comparison with the $\Omega$-effect, as $C_\alpha<<C_\Omega$. This assumption seems to be supported by energy transfers study in decadal magnetic cycles of global MHD simulations (see e.g., Figure 23. of \citealt{Brun+2021}). However, the $\alpha^2\Omega$ model may still be relevant as the $\alpha$-effect operates in a very thin layer such as the tachocline (see \citealt{Charbonneau2020LRSP} and references therein). We tried to apply such an $\alpha^2\Omega$ model in some of our runs (not shown here) but did not find significant differences from a dynamical point of view. Nevertheless, we did not do it through the entire parameters space that we explored in this paper.\\
We note here that as we chose $C_\Omega>>C_\alpha$ or $C_S$ for this study, therefore the toroidal magnetic energy is always higher than the poloidal magnetic energy. The ratio of these two energies can vary by two orders of magnitude for a given dynamo type, when modifying the $\alpha$-effect location. The models considered in this work produce a steady-state energy ratio between $1.8\times10^2$ and $2.4\times10^5$.
\vspace{1\baselineskip}\\
Using a different anti-solar DR profile that does not preserve the same shear properties applied in the present study, \citet{karak_stellar_2020} performed a study using $\alpha^2\Omega$ mean-field dynamos.
Symmetric and antisymmetric $\alpha$-profiles were considered, adding anisotropy for some models. We note that these authors only considered profiles lying mainly in the bulk of the CZ, hence did not use negative $\alpha$-effect values. They conclude that global magnetic polarity reversals are possible for anti-solar DR, but only if a quenching of the DR shear by the magnetic field is prescribed.\\
We show here that global polarity reversals for anti-solar DR are possible without any retro-action on the part of the magnetic field on the large-scale flows by exploring dynamo mechanisms with deeper location than what was done before. However, we could extend our model out of the kinematic regime by taking into account this retro-action. This could be done in a future work for instance with the Malkus-Proctor effect \citep{M-P1975}, characterizing the macroscopic feedback by the Lorentz force on the flows. Indeed, recent studies like \cite{strugarekReconcilingSolarStellar2017} and \cite{Brun+2021} underline the major role of this feedback mechanism in dynamics of decadal solar-type magnetic cycles.\\
Non-linear feedback also occurs at small scale in dynamo action, and this aspect has been modeled as a simple alpha quenching in this work. Nevertheless, recent studies underline that mean-field coefficients can have non-trivial dependencies, for instance, in high Rossby regime where anti-solar DR are expected \citep[see][and reference therein]{warneckeInvestigatingGlobalConvective2021}. Such complex feedbacks are beyond the scope of the present study and shall be considered in future works (see also \citealt{pipinSolarDynamoCycle2021}).

\section{Conclusions}\label{sec:ccl}
In this work, we explore how the dynamo mechanism behaves under different physical prescriptions of the mean-field theory. Our main interest is to assess whether magnetic cycles could be sustained in an anti-solar differential rotation regime. We consider $\alpha\Omega$ and Babcock-Leighton models of solar-type stars to this end.\\
We first set up the two types of models, so that they may reproduce the well-known characteristics of the dynamo in its solar context, based on the exploration of the models by the community over the past half-a-century. We obtained similar surface dynamics representative of those at work in our star: a cycle with polarity reversal, a migration of the activity towards the equator as well as the diffusion of the opposite polarity towards the poles during a cycle. We also found the expected characteristics in each hemisphere at the bottom of the CZ for $\alpha\Omega$ and BL models, namely, Parker-Yoshimura waves for the former and strong equatorward migrating branches for the latter.
The dynamo is, of course, closed by an $\Omega$-effect induced by a solar DR profile.\\
Reversing this DR profile toward an anti-solar regime, we studied its effect on the dynamo mechanism. With a simple $\alpha\Omega$ dynamo model, we show that magnetic cycles can occur in anti-solar DR regime with an $\alpha$-effect saddled on the tachocline at the base of the CZ. This result holds regardless of the $\alpha$-effect sign considered in the northern hemisphere (see mirror solutions when inverting the sign of $C_\alpha$), which seems to contradict what has been postulated in previous studies as \cite{karak_stellar_2020} that have mainly been focused on $\alpha$-profiles distributed throughout the bulk of the CZ. Additionally, they need to prescribe a magnetic feedback in order to get global polarity reversals in anti-solar DR regime, which is not the case for the present study. Finally, we observed an increase of the cycle frequency, which is nearly doubled, due to the emergence of a strong quadrupolar component of the magnetic topology in this model.\\
In contrast, we show the disappearance of the cyclic behavior, for positive $\alpha$-effect in the CZ and BL model, once an anti-solar DR regime is applied. We propose in Figure \ref{fig:GraphAOBLcomplet} a geometrical interpretation, relying on the fact that Coriolis effect in anti-solar DR could lead to polar flux addition, instead of cancellation, in the well-known solar DR regime. This could explain why no global polarity reversals are observed in most of anti-solar DR cases.
In the solar DR regime, we found that the magnetic cycle period in $\alpha\Omega$ dynamos depends on the location and thickness of our $\alpha$-source term and it is generally longer for deeper locations and broader source terms. Indeed, the closer to the surface the $\alpha$-effect, the less magnetic flux will have to be cancelled by the next polarity in the cycle; hence, the cycle period is shorter.\\
In the anti-solar DR regime, we found that moving the $\alpha$-effect away from the tachocline is enough to suppress the magnetic cycle in such model; namely, if we segregate spatially both the poloidal and toroidal source terms. The disappearance of cyclic activity in BL models under anti-solar DR regime is therefore likely to be caused by the position of its poloidal source term, rather than its non-local nature.\\
We further confirmed this finding by studying the influence of MC. We concluded for all cases that the MC generally does not help to obtain a magnetic cycle if one did not exist before. This was also found in \citet{Bonanno2002} for $\alpha^2\Omega$ models. For solar cases, it even sometimes leads to suppress the cyclical behavior of some models which initially had it, especially when the poloidal generation is located near the surface and the MC (via $R_{\rm m}$) is relatively fast.\\
Finally, we completed this investigation by considering the opposite side of the problem and trying to make the reference anti-solar BL model cyclic. To this end, we tried to modify the different parameters of the model, but the dynamo remained stationary in all cases . The only way we found it possible to make a BL model with anti-solar DR cyclic was to consider an additional $\alpha$-effect at the base of the CZ. In that case, the dynamo is eventually cyclic, albeit with a polarity reversal occurring only locally at mid and low latitudes.\\
In conclusion, we found that the BL model alone, in its simplest expression and in the anti-solar DR regime, seems to prevent any form of cyclic stellar activity. This is mainly due to a flux addition, illustrated in Figure \ref{fig:GraphAOBLcomplet} (also proposed by \citealt{karak_stellar_2020} in their Figure 1). We do not find a significant drop of the large-scale magnetic field topology for anti-solar DR regime that could explain a possible break of gyrochronology for slow rotators, as proposed by \cite{metcalfeMagneticEvolutionDisappearance2017}. Likewise, the $\alpha\Omega$ model generally do not produce cycle in the anti-solar DR regime, except when the $\alpha$-effect is localized at the tachocline. The key point of such models is then the overlap between the poloidal generation through $\alpha$-effect and the toroidal generation through radial shear. In these models we also do not observed any drop of the large-scale field. Observational constraints on the magnetism of stars with anti-solar DR would therefore provide a fantastic constraint on dynamo acting within cool stars.\vspace{1\baselineskip}\\
We thank the anonymous referee for constructive comments that have led to the improvement of the manuscript. We also acknowledge financial support by ERC Whole Sun Synergy grant \#810218., funding by INSU/PNST grant, CNES/Solar Orbiter and CNES/PLATO funds.

\bibliographystyle{aa}
\bibliography{AntiSolPap}

\begin{appendix}
\section{Fine-tuned $\alpha-\Omega$ solar model}
\begin{figure}[ht!]
    \centering
    \includegraphics[width=\linewidth]{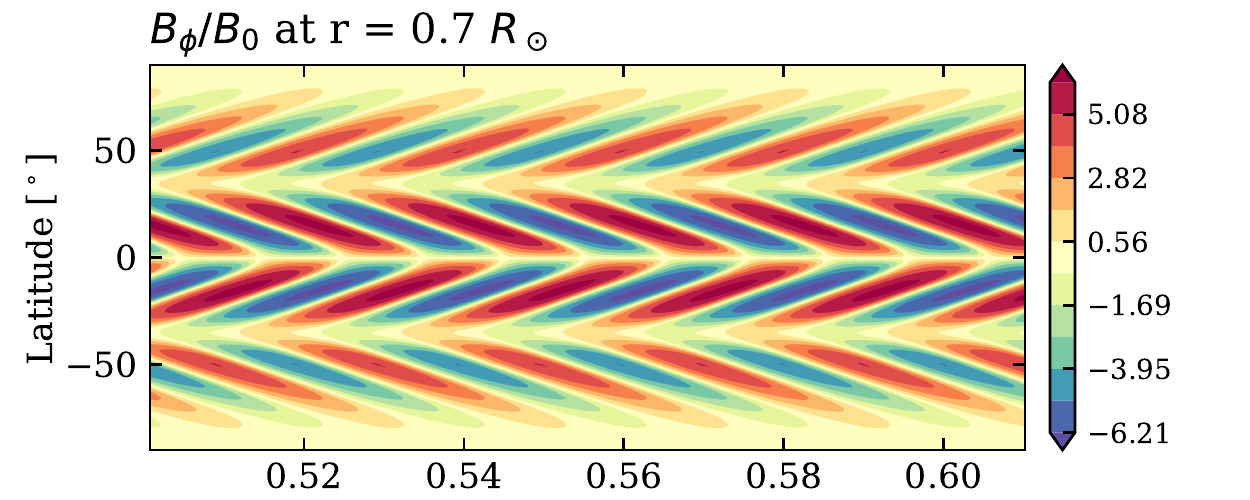}
    \includegraphics[width=\linewidth]{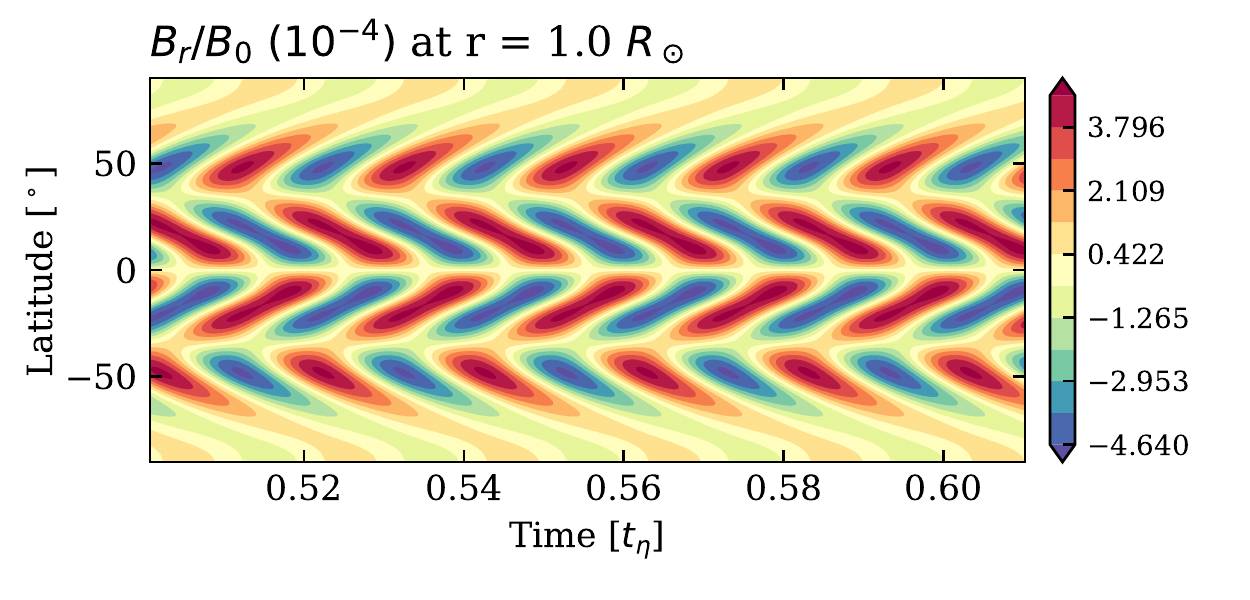}
    \caption{Butterfly diagram of the reference $\alpha\Omega$ model in solar DR regime of Figure \ref{fig:Sun_AO}, fine-tuned in order to get our Sun magnetic cycle period with $C_\Omega=9.3\times10^4$. We recall that for this specific case, 22 years correspond to 0.021 $t_\eta$.}
    \label{fig:SunModelAOtuned}
\end{figure}
\section{Meridional evolution of cyclic reference models}
\begin{figure*}
    \centering
    \includegraphics[width=0.24\linewidth]{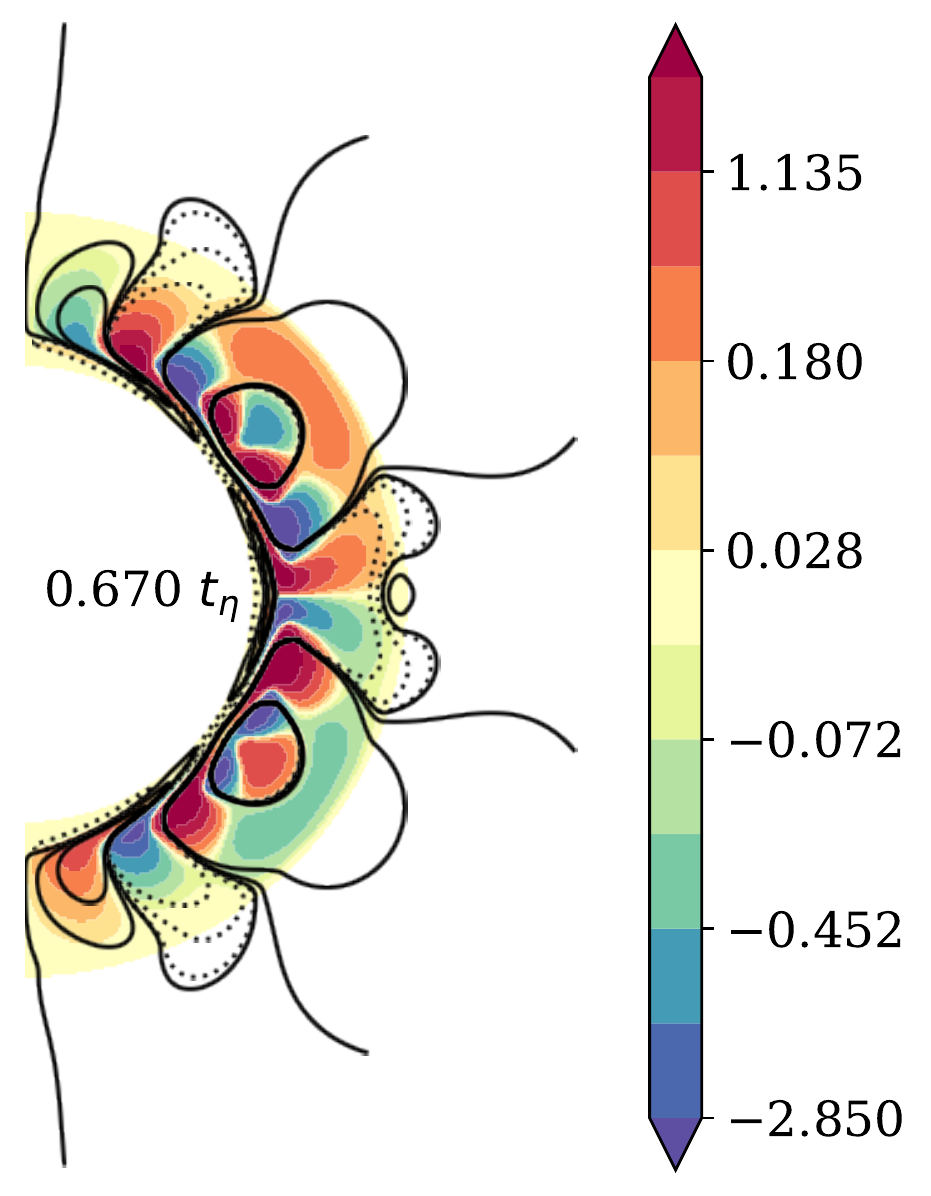}
    \includegraphics[width=0.24\linewidth]{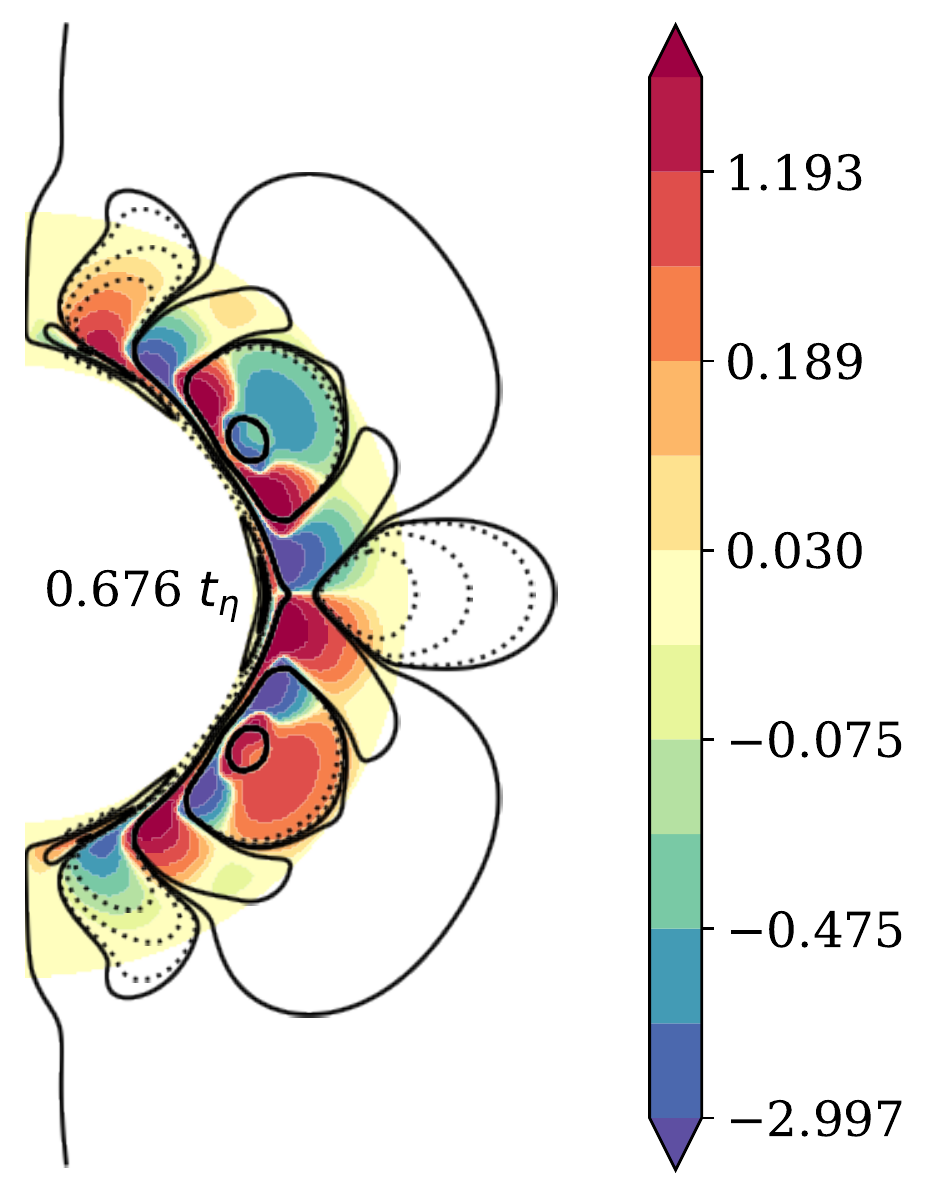}
    \includegraphics[width=0.24\linewidth]{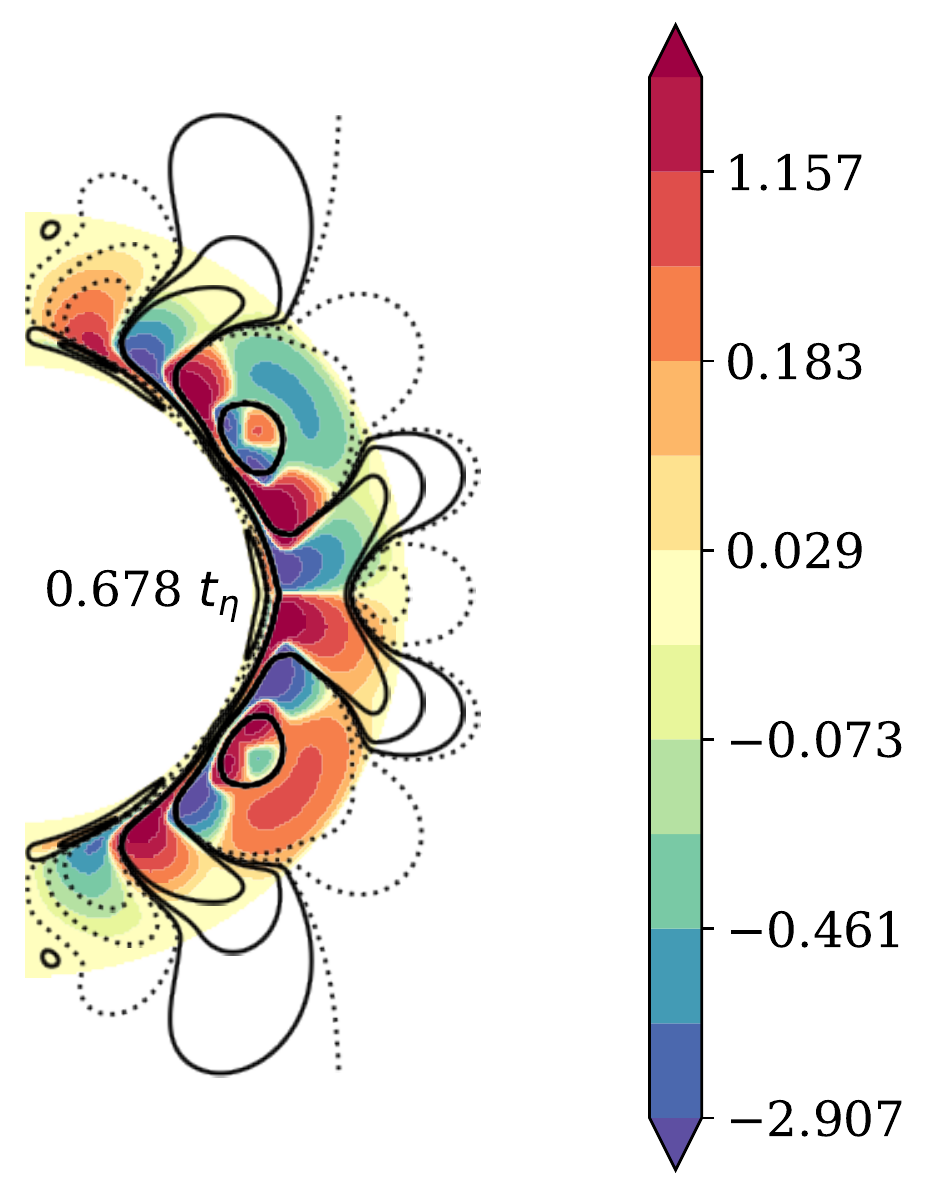}
    \includegraphics[width=0.24\linewidth]{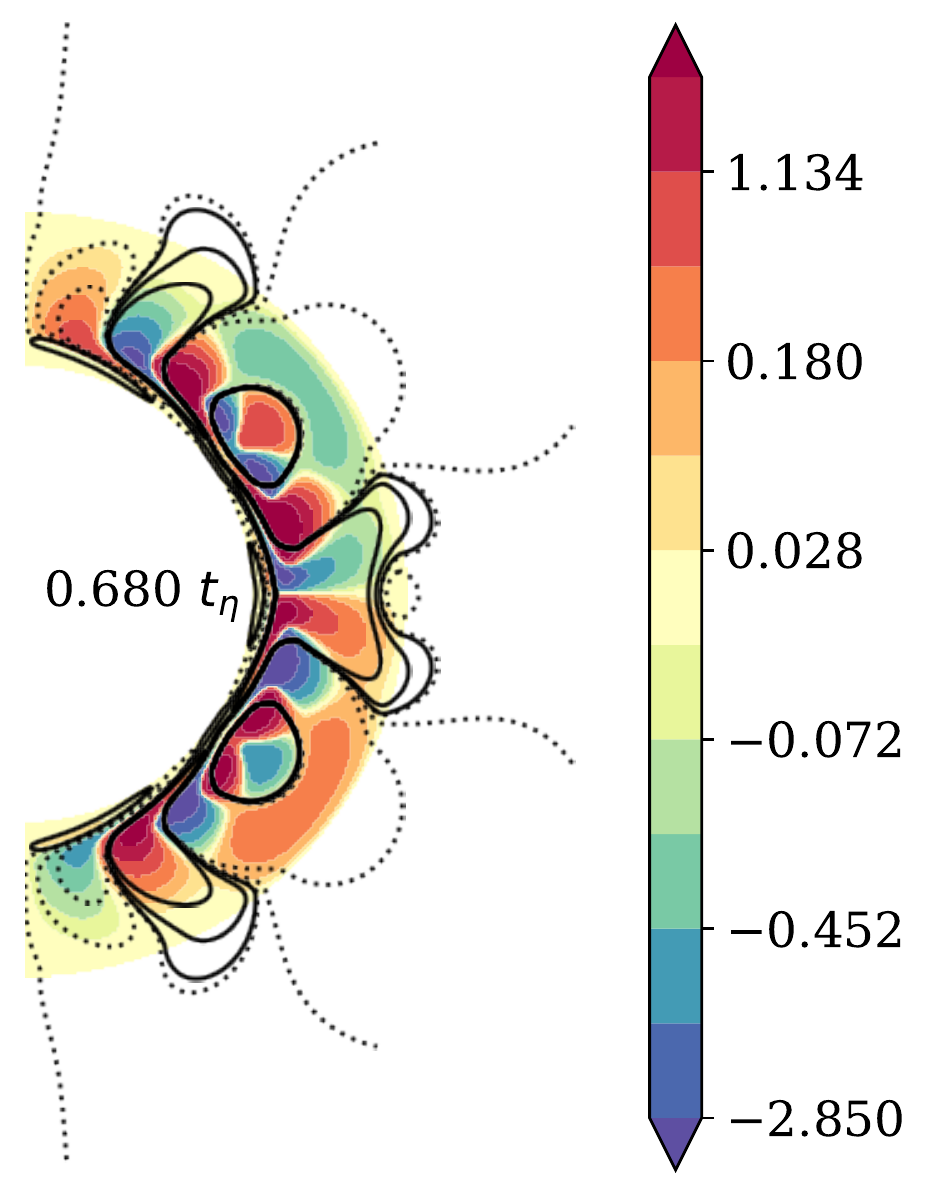}
    \includegraphics[width=0.24\linewidth]{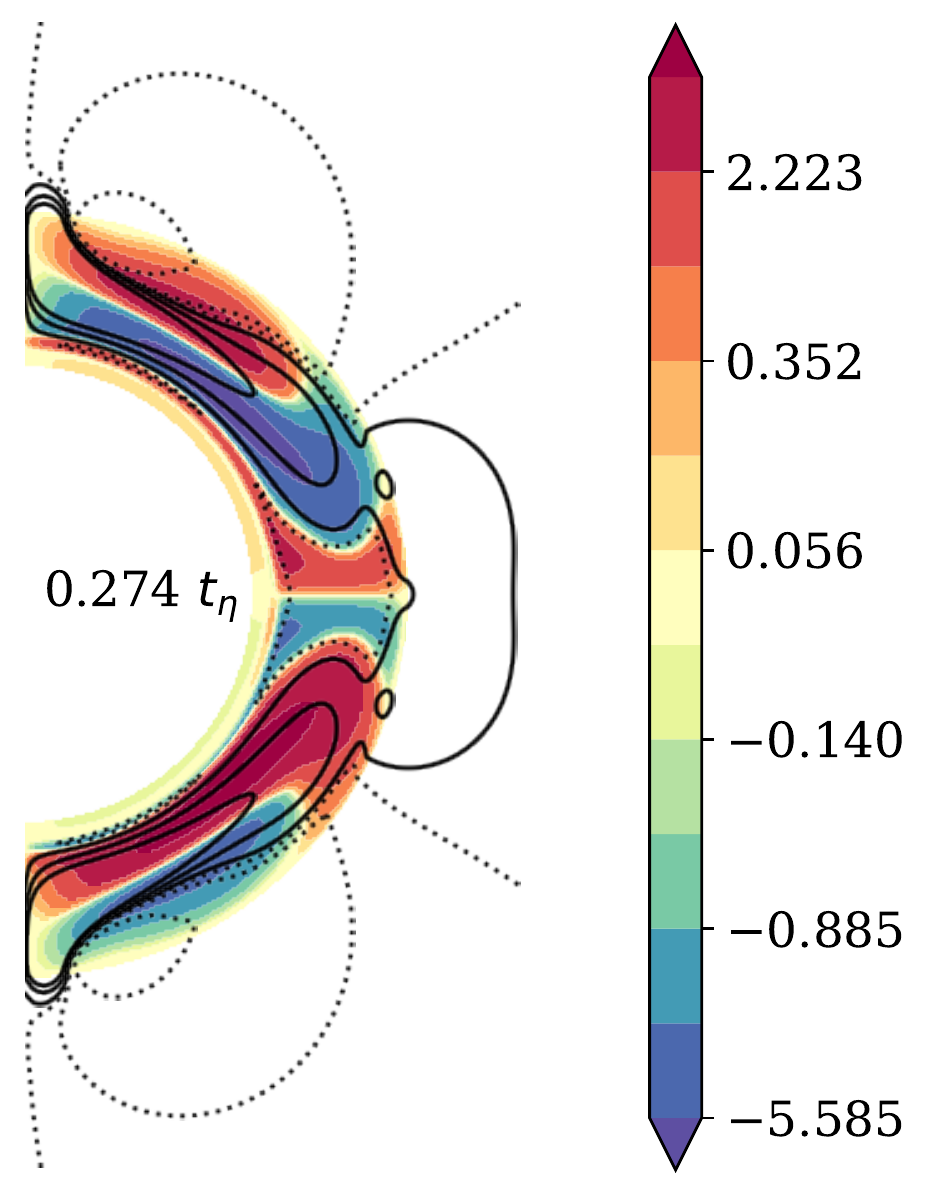}
    \includegraphics[width=0.24\linewidth]{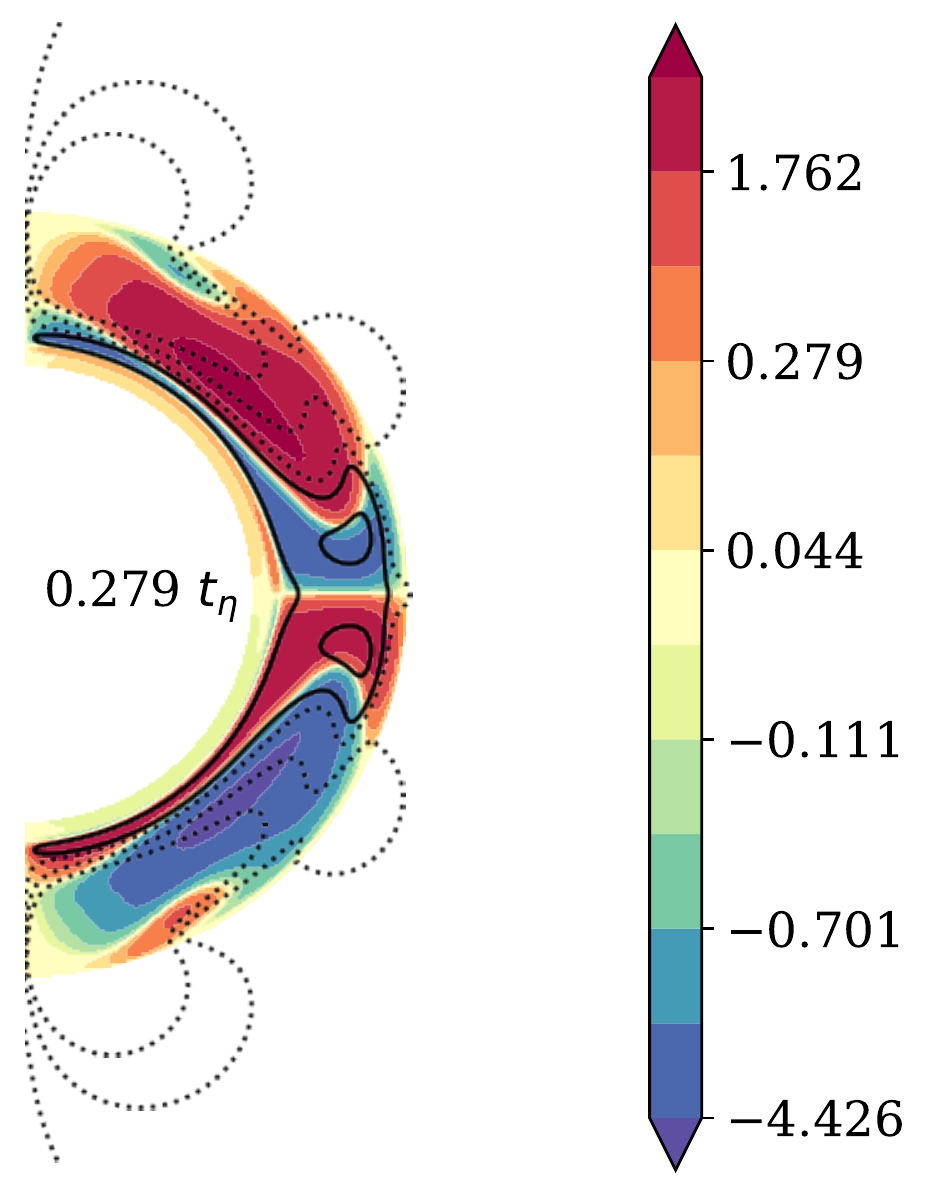}
    \includegraphics[width=0.24\linewidth]{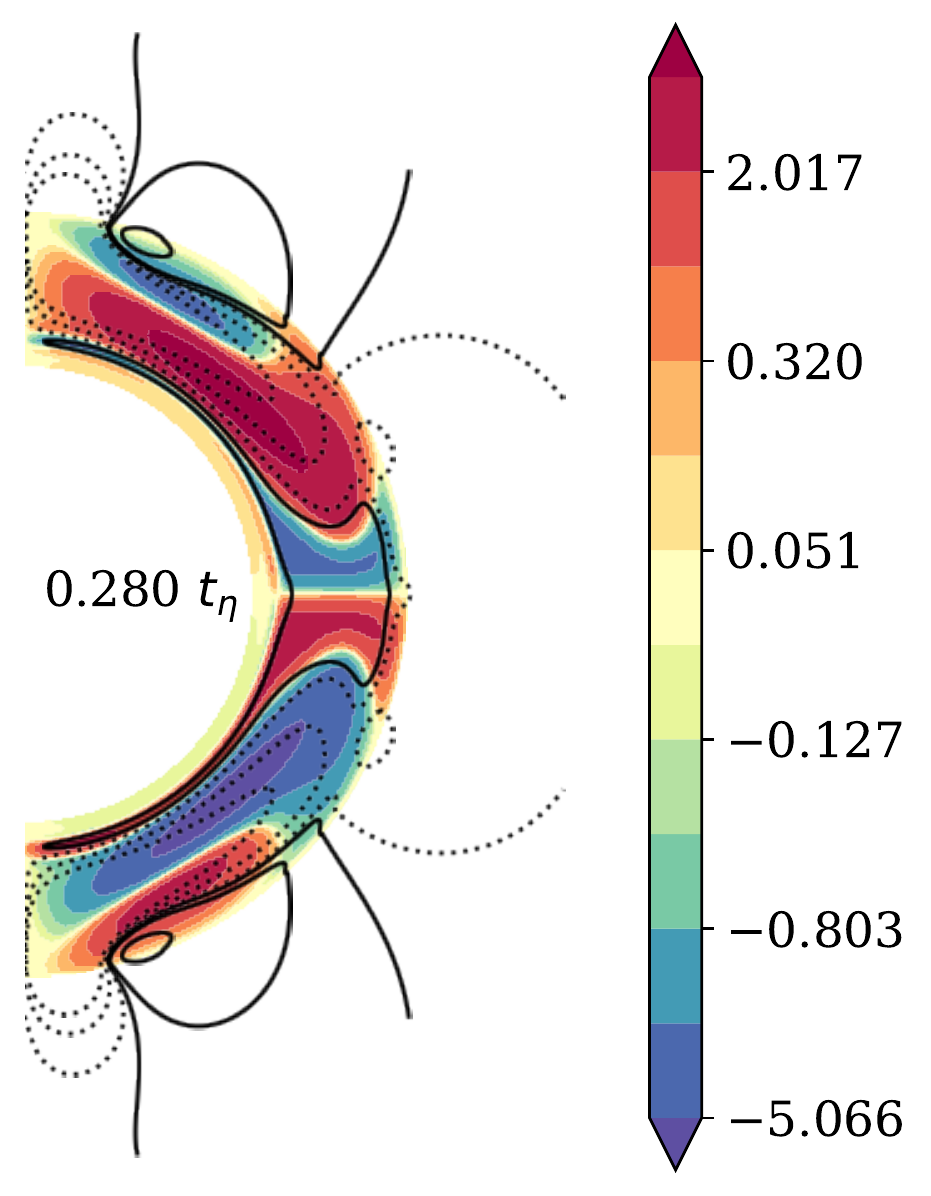}
    \includegraphics[width=0.24\linewidth]{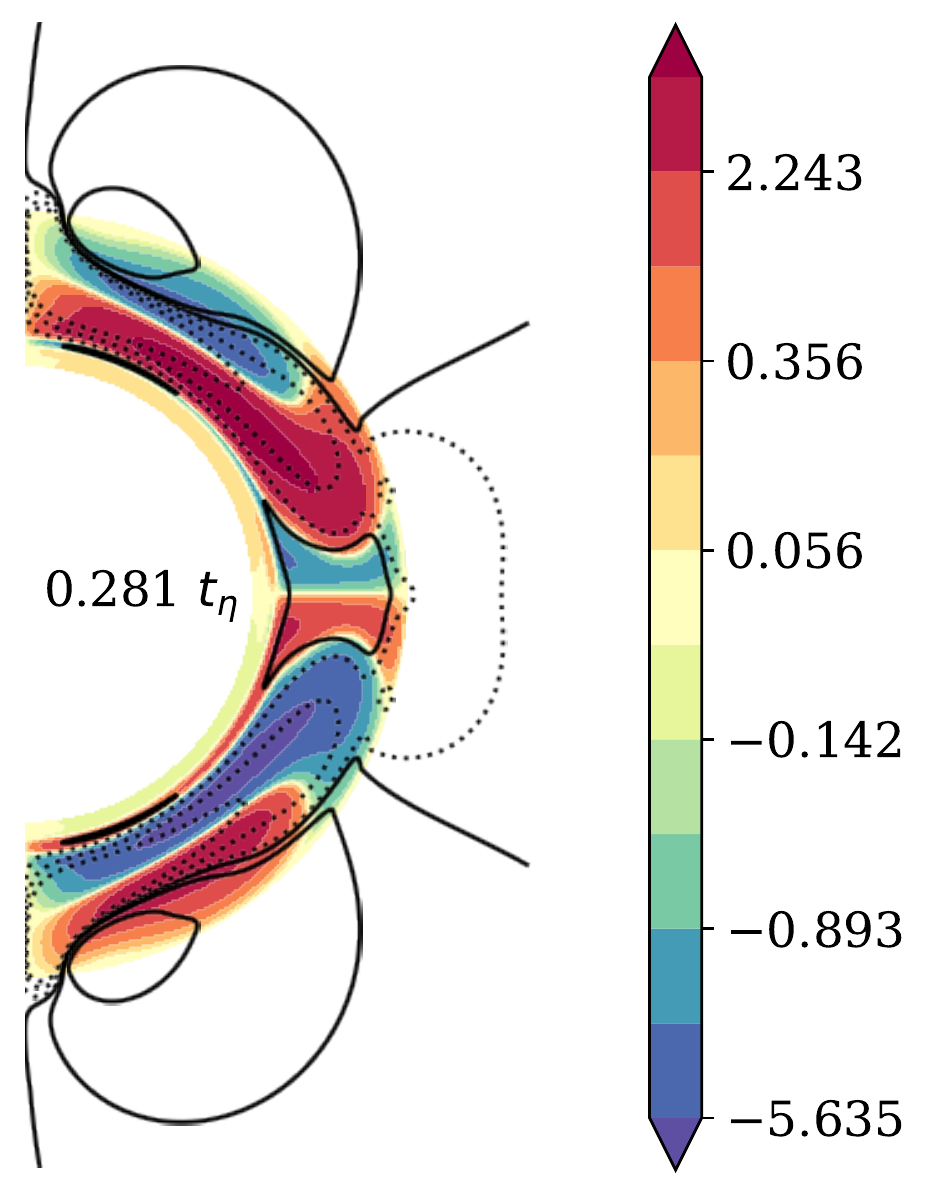}
    \includegraphics[width=0.24\linewidth]{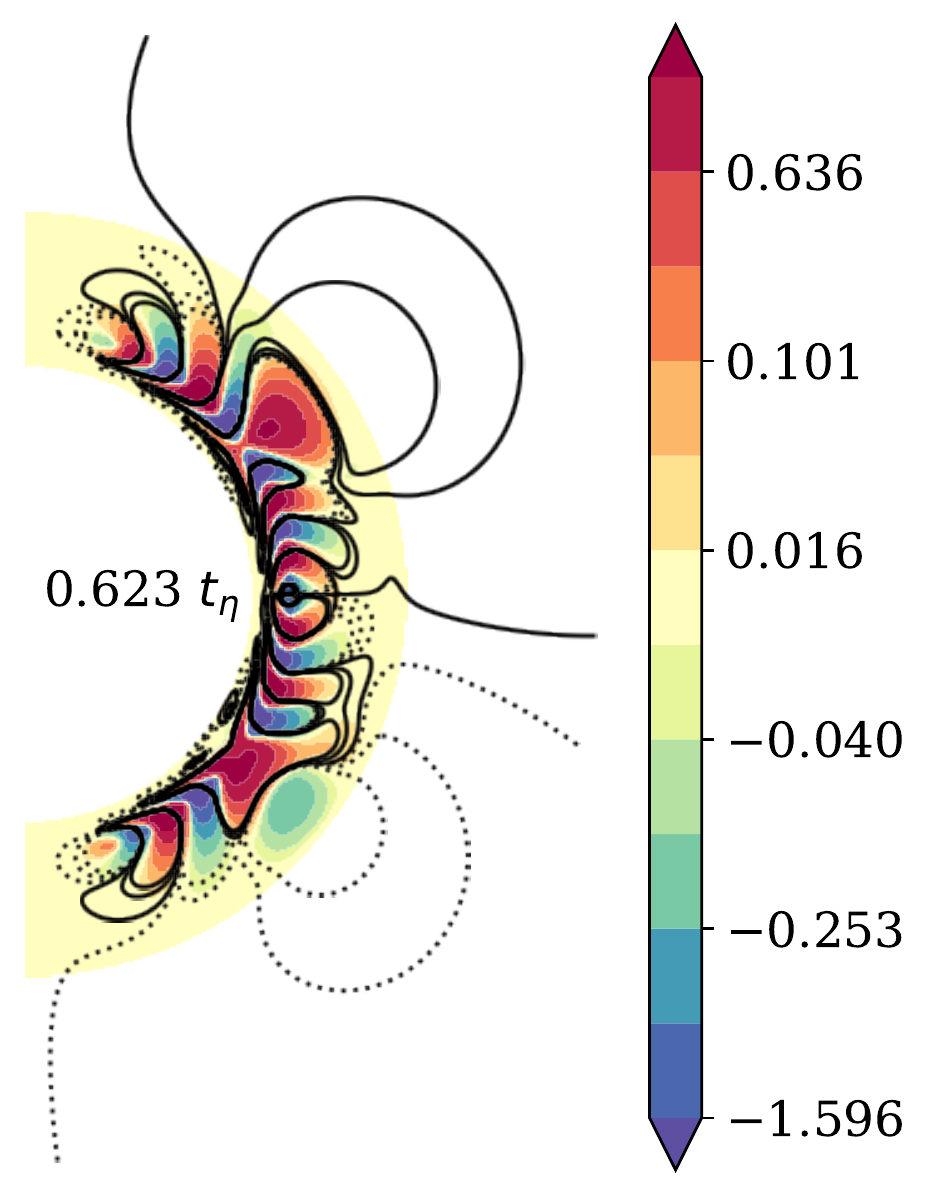}
    \includegraphics[width=0.24\linewidth]{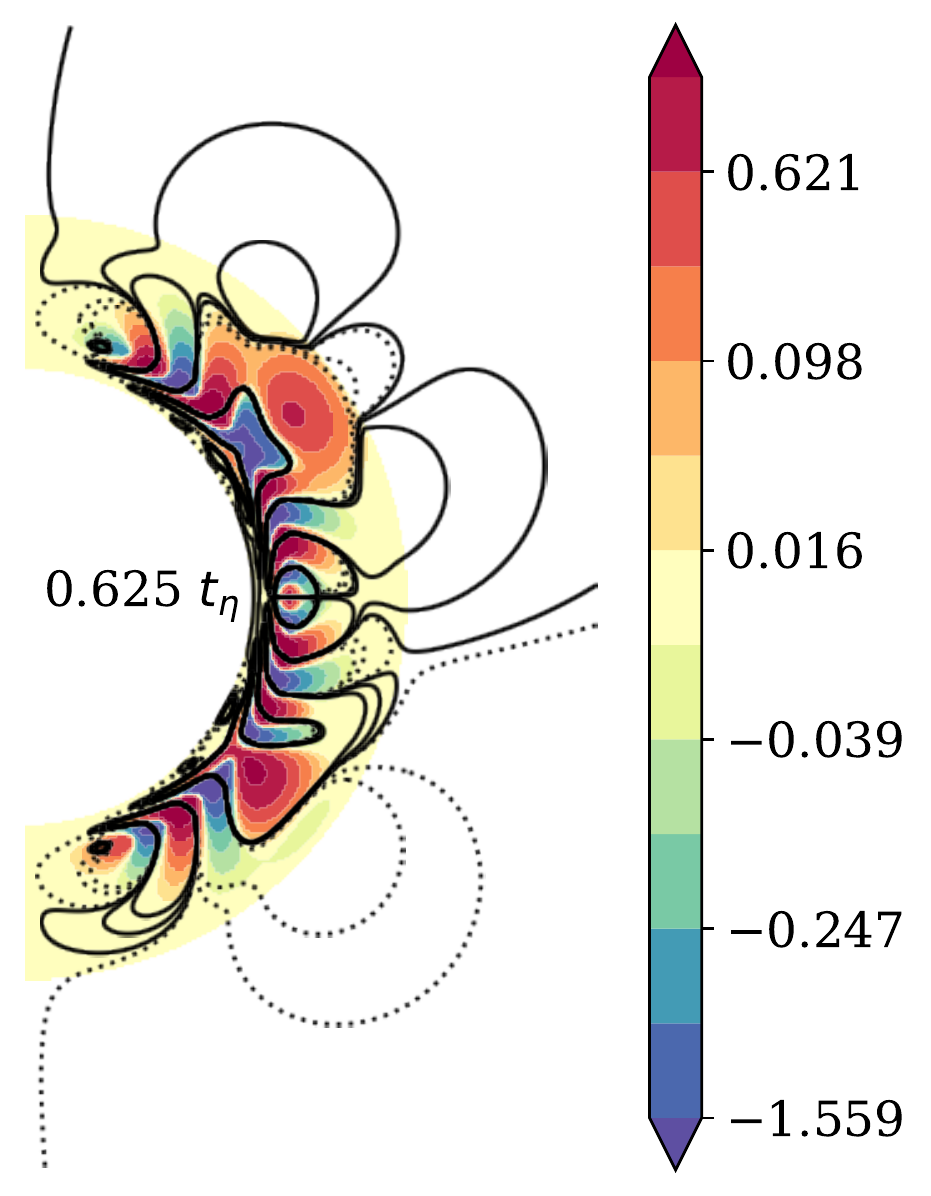}
    \includegraphics[width=0.24\linewidth]{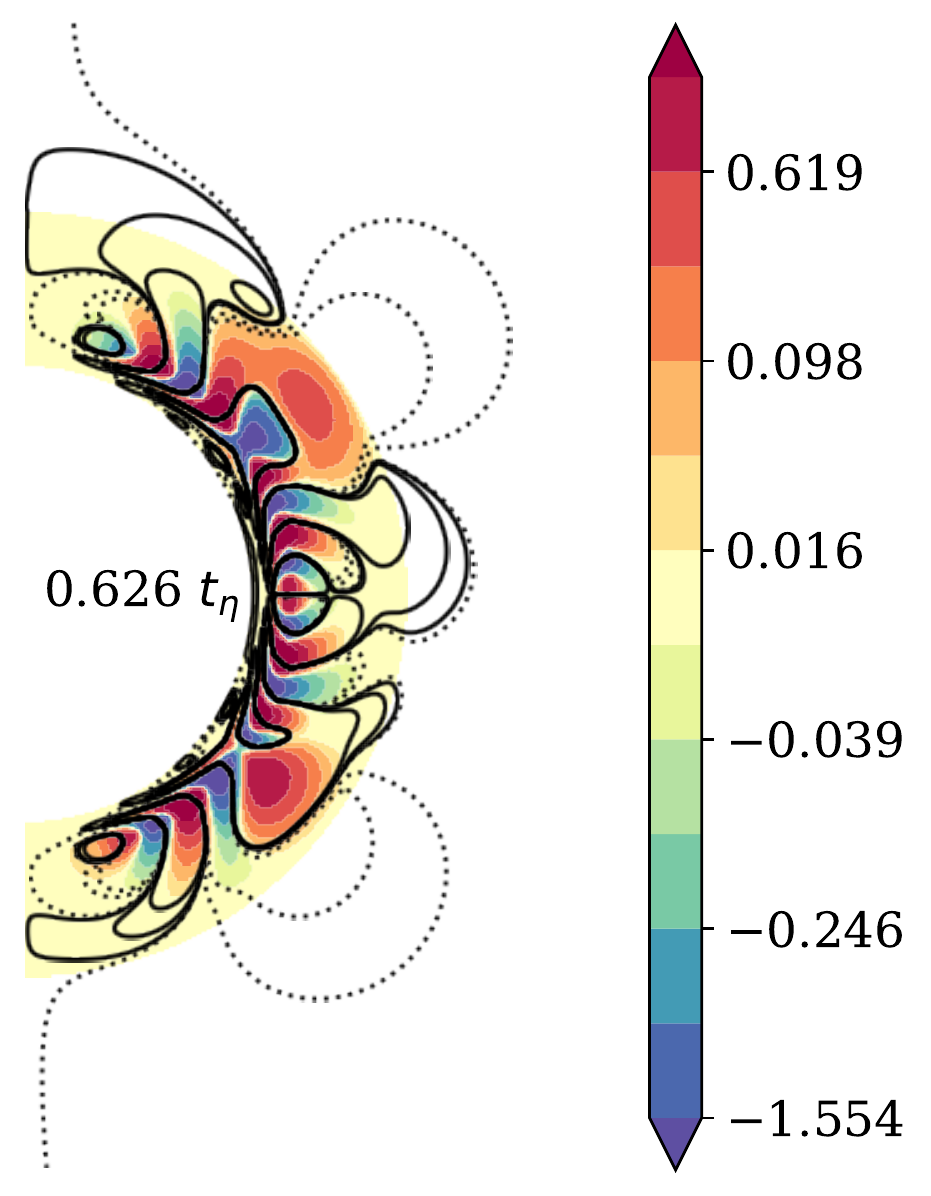}
    \includegraphics[width=0.24\linewidth]{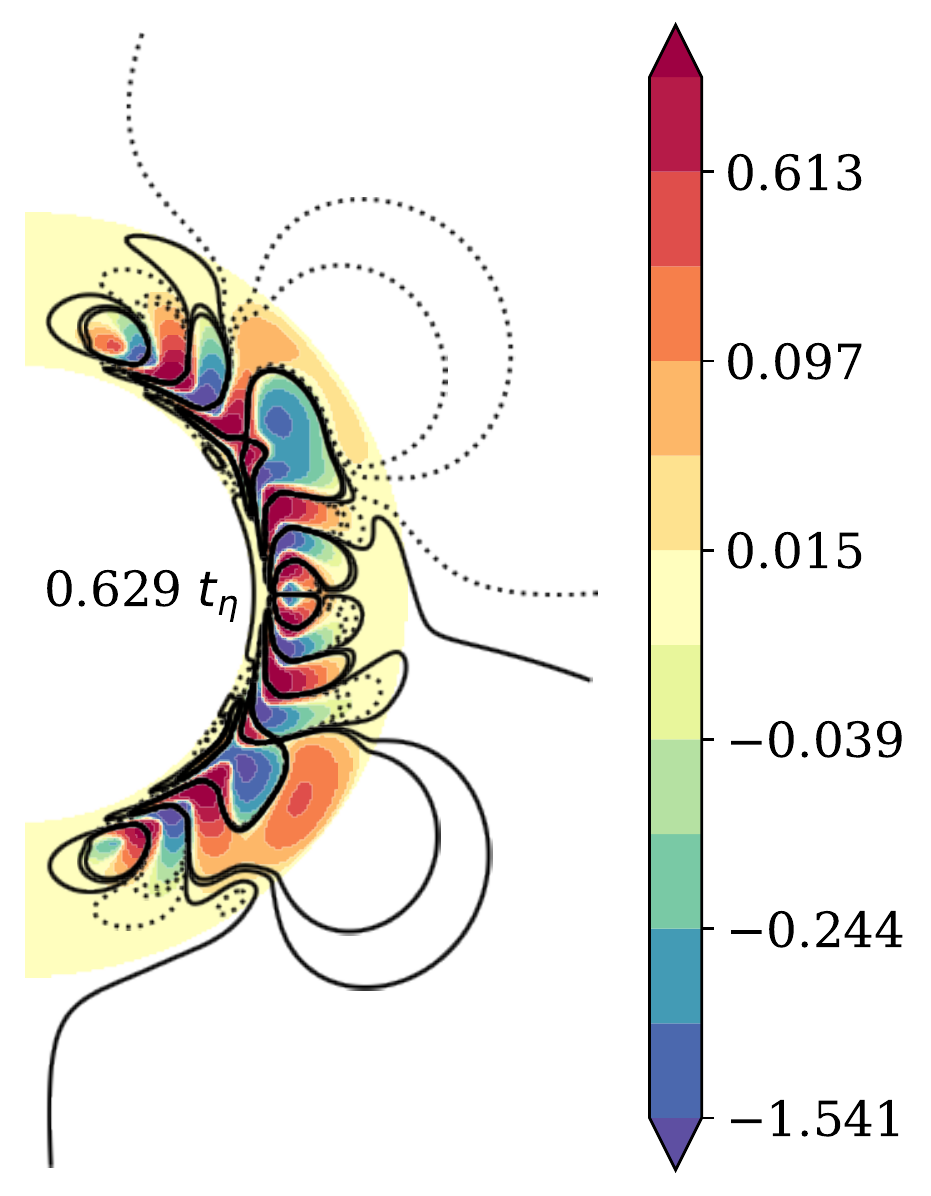}
    \caption{Meridional cut from 0.6 to 1 $R_\odot$, of magnetic components for three reference models of Section \ref{sec:DR_influence}: Solar DR $\alpha\Omega$ dynamo (first line), the solar DR BL dynamo (second line) and the anti-solar DR $\alpha\Omega$ dynamo (third line). Colors are mapped on a logarithmic scale for $B_\phi/B_0$. Black continuous contours are positive $A_\phi$ iso-contours while dotted lines are for negative contours, i.e., $B_{pol}$ clockwise oriented field lines, and anti-clockwise respectively. Field lines above the surface r = 1$R_\odot$ result from a potential extrapolation. Four different time steps are proposed for each model. We recall that in this study 22 years correspond to 0.014 $t_\eta$.}
    \label{fig:MeridCuts}
\end{figure*}
\end{appendix}
\end{document}